\documentclass[reprint,
superscriptaddress,
amsmath,amssymb,
aps,
prl,
]{revtex4-2}

\usepackage{graphicx}
\usepackage{dcolumn}
\usepackage{bm}
\usepackage{extarrows}
\usepackage{dsfont}
\usepackage{physics}

\begin{document}


\title{Microscopic Theory of Vibrational Polariton Chemistry}
 
\author{Wenxiang Ying}
\affiliation{Department of Chemistry, University of Rochester, Rochester, NY 14627, USA}
\author{Michael A.D. Taylor}
\affiliation{The Institute of Optics, Hajim School of Engineering, University of Rochester, Rochester, NY 14627, USA}
\author{Pengfei Huo}
\email{pengfei.huo@rochester.edu}
\affiliation{Department of Chemistry, University of Rochester, Rochester, NY 14627, USA}
\affiliation{The Institute of Optics, Hajim School of Engineering, University of Rochester, Rochester, NY 14627, USA}


\begin{abstract}
We present a microscopic theory that aims to explain the vibrational strong coupling (VSC) modified reaction rate constant. The analytic theory is based on a mechanistic conjecture that cavity modes promote the transition from the ground state to the vibrational excited state of the reactant, which is the rate-limiting step of the reaction. The theory explains the observed resonance effect at the normal incident angle. Assuming the coherent vibrational energy transfer picture,  the theory can also explain the collective effect and makes several predictions that are experimentally verifiable. 
\end{abstract}

\maketitle
Recent experiments~\cite{Ebbesen_angew_2016, Ebbesen_angew_2019_2, Ebbesen_science_2019, Ebbesen_nanophotonics_2020, Ebbesen_angew_2019, Lather_2022} have demonstrated that chemical reaction rate constants can be suppressed~\cite{Ebbesen_angew_2016, Ebbesen_angew_2019_2, Ebbesen_science_2019, Ebbesen_nanophotonics_2020, Hirai2020-Prins, Simpkins2023} or enhanced~\cite{Ebbesen_angew_2019, Lather2020JPCL, Lather_2022} by resonantly coupling molecular vibrations to quantized radiation modes inside a Fabry-P\'erot (FP) microcavity~\cite{Hirai2020C, Nagarajan2021JACS, Simpkins-review}. This effect has the potential to selectively slow down competing reactions~\cite{Ebbesen_science_2019} or speed up a target reaction, thus achieving mode selectivities and offering a paradigm shift in chemistry. Despite extensive theoretical efforts~\cite{Horwell_2023_review, Galego2019, Ribeiro2019, Nitzan2019, Vurgaftman2020, TaoJCP2020, Zhdanov2020, Jorge2020, Li_2021, Rubio_2021, Li2021JPCL,Li2021ACIE, Litao_jcp_2022, Arkajit_JCP_2022, Joel_prl_2022, Dou_2022, Wang2022JPCL, Wang2022Collective, Sun2022JPCL, Fischer2022JCP, Arkajit_jpcl2022, Mondal2022JCP, Cao_2022, Kansanen2023, YZ2023PGH, Arkajit_2022,Limmer_2023,Simpkins2023}, the fundamental mechanism and theoretical understanding of the cavity-modified ground-state chemical kinetics remain elusive~\cite{Horwell_2023_review,Wang2021AP,sidler2022JCP, Mandal2022C}. To the best of our knowledge, there is no unified theory that can explain all of the observed phenomena in the vibrational strong coupling (VSC) experiments~\cite{Horwell_2023_review}, including (1) The resonant effect, which happens when the cavity frequency matches the bond vibrational frequency, $\omega_\mathrm{c}=\omega_{0}$, but also only happens when  the in-plane photon momentum is $k_{\parallel} = 0$, (2) The collective effect, which is the increase in the {\it magnitude} of the VSC modification when increasing the numbers of molecules $N$ (or concentration $N/\mathcal{V}$)~\cite{Ebbesen_angew_2019, Ebbesen_angew_2016, Ebbesen_nanophotonics_2020}, (3) The process is driven by thermal fluctuations without optical pumping~\cite{Ebbesen_angew_2016, Ebbesen_science_2019}. (4) The dipoles are assumed to have an isotropic disorder~\cite{Horwell_2023_review}.

 We aim to develop a microscopic theory to explain all of these observed VSC effects. Let us consider $N$ identical molecules coupled to many radiation modes inside a Fabry-P\'erot cavity,
\begin{align} \label{Ham}
    \hat{H}&=\sum_{j=1}^{N}\frac{\hat{P}_{j}^2}{2M} + V(\hat{R}_{j})+\hat{H}_{\nu}+ \hat{H}_\mathrm{loss} (\hat{q}_{\bf k},\hat{x}_{{\bf k},\zeta})\\
    &+\sum_{\bf k}\frac{\hat{p}^2_{\bf k}}{2} + \frac{\omega^2_{\bf k}}{2} \Big(\hat{q}_{\bf k} + {\frac{\lambda_\mathrm{c}}{\omega_{\bf k}}} \cdot \sum_{j=1}^{N}{\mu}(\hat{R}_{j})\cdot\cos\varphi_{j} \Big)^2,\notag
\end{align}
where $\hat{R}_{j}$ is the reaction coordinate for the $j_\mathrm{th}$ molecule, $V(\hat{R})$ is the ground state potential for all reaction molecules (typically double well potential), and $\mu(\hat{R}_{j})$ is the dipole associated with the ground electronic state (electronic permanent dipole). The Fabry-P\'erot cavity has the following dispersion relation
\begin{align}\label{eq:FP-freq}
\omega_{\bf k}(k_{\parallel}) = \frac{c}{n_\mathrm{c}}\sqrt{k^2_{\perp} + k^2_{\parallel}} = \frac{ck_{\perp}}{n_\mathrm{c}}\sqrt{1 + \tan^2 \theta},
\end{align}
where $c$ is the speed of light, $n_\mathrm{c}$ is the refractive index of the cavity, and $\theta$ (usually referred to as the incident angle) is the angle of the photonic mode wavevector, $\mathbf{k}$, relative to the norm direction of the mirrors. When $k_{\parallel}=0$ (or $\theta = 0$), the photon frequency is
\begin{equation} \label{wc_def}
    \omega_\mathrm{c}\equiv\omega_{{\bf k}}(k_{\parallel}=0)=ck_{\perp} / n_\mathrm{c}.
\end{equation}
The cavity frequency $\omega_{\bf k}$ in Eq.~\ref{Ham} is associated with the wavevector ${\bf k}$, according to Eq.~\ref{eq:FP-freq}, where $\hat{q}_{\bf k} = \sqrt{\hbar/(2\omega_{\bf k})}(\hat{a}_{\bf k}^{\dagger} + \hat{a}_{\bf k})$ and $\hat{p}_{\bf k} = i\sqrt{\hbar\omega_{\bf k}/2}(\hat{a}_{\bf k}^{\dagger} - \hat{a}_{\bf k})$, $\hat{a}_{\bf k}$ and $\hat{a}_{\bf k}^{\dagger}$ are the photonic field annilation and creation operators for mode ${\bf k}$, respectively. The light-matter coupling strength is $\lambda_\mathrm{c} = \sqrt{1/(\epsilon_0 \mathcal{V})}$, where $\epsilon_0$ is the effective permittivity inside the cavity and $\mathcal{V}$ is the cavity quantization volume. Each reaction coordinate $R_{j}$ is coupled to its own local phonon bath described by $\hat{H}_{\nu}$. Each cavity mode $\hat{q}_{\bf k}$ couples to its independent bath $\{\hat{x}_{{\bf k},\zeta}\}$, accounting for the cavity loss. The cavity modes $\{\hat{q}_{\bf k}\}$ couple to the dipole of each molecule $\mu(\hat{R}_{j})$, where $\varphi_{j}$ is the relative angle between the dipole vector and the field polarization direction. Details of the Hamiltonian are provided in Supplemental Material~\cite{SI}, Sec. I.

Let us consider an idealized reaction mechanism outside the cavity as $|\nu_\mathrm{L}\rangle \xlongrightarrow{\text{$k_1$}} |\nu'_\mathrm{L}\rangle \xlongrightarrow{\text{$k_2$}} |\nu'_\mathrm{R}\rangle \xlongrightarrow{\text{$k_3$}} |\nu_\mathrm{R}\rangle$,
where $|\nu_\mathrm{L}\rangle$ denotes the vibrational ground state of the reactant (left well), $|\nu'_\mathrm{L}\rangle$ denotes the vibrationally excited state of the reactant, and similar for the product (right well). See Fig.~S1 of the Supplemental Material~\cite{SI}. These vibrational diabatic states can be directly obtained by computing the vibrational eigenspectum of $V(\hat{R})$ and then diabatizing them. The simplified mechanism for the reaction is that the thermal activation process causes the transition of $|\nu_\mathrm{L}\rangle \to |\nu'_\mathrm{L}\rangle$. Then the reaction occurs through the diabatic couplings between $|\nu'_\mathrm{L}\rangle$ and $|\nu'_\mathrm{R}\rangle$, followed by a vibrational relaxation of the product state, $|\nu_\mathrm{R}\rangle$. The symmetric double well model~\cite{Arkajit_2022} is used for the model reaction, with details in Supplemental Material~\cite{SI}, Sec II. The rate-limiting step for the entire process is $k_1$, where $k_{2}\gg k_{1}$ such that the population of $|\nu'_\mathrm{L}\rangle$ and $|\nu'_\mathrm{R}\rangle$ reaches a steady state (plateau in time), and from the steady-state approximation, the overall rate constant for the reaction is $k_{0}\approx k_{1}$. 

Considering many molecules, we focus on the single excitation subspace. This includes the ground state $|G\rangle$ and $N$ singly excited states $|\nu_{j}\rangle$ (where $j\in[1,N]$ labels the molecules), defined as  
\begin{subequations}
\begin{align}&|G\rangle\equiv|\nu^{1}_\mathrm{L}\rangle...\otimes...|\nu^{j}_\mathrm{L}\rangle\otimes...|\nu^{N}_\mathrm{L}\rangle,\\&|\nu_{j}\rangle\equiv|\nu^{1}_\mathrm{L}\rangle...\otimes...|\nu'^{j}_\mathrm{L}\rangle\otimes...|\nu^{N}_\mathrm{L}\rangle.
\end{align}
\end{subequations}
The vibrational transition dipole matrix element is 
\begin{equation}
    \mu_{\mathrm{L}\mathrm{L}'}=\langle \nu'^{j}_\mathrm{L}|\mu(\hat{R}_{j})|\nu^{j}_\mathrm{L}\rangle,
\end{equation}
which is identical for all molecule $j$. When measuring the absorption spectra of the molecule, the optical response shows a peak at the quantum vibrational frequency $\omega_{0}=(E_{\nu'_\mathrm{L}}-E_{\nu_\mathrm{L}})/\hbar$. 

In the singly excited manifold, the collective ``bright state"  is  defined as $|\mathrm{B}\rangle=\frac{1}{\sqrt{N}}\sum_{k=1}^{N} |\nu_{k}\rangle$. The light-matter coupling term $\propto \sum_{{\bf k},j}\hat{q}_{\bf k}\otimes\mu(\hat{R_{j}})$ in Eq.~\ref{Ham} will hybridize bright states and photon-dressed ground states, generating polariton states~\cite{Hopfield1958PR}. When all dipoles are fully aligned, such that $\cos\varphi_{j}=1$, and under the resonant condition $\omega_{\bf k}(k_{\parallel})=\omega_{0}$, the light-matter hybridization generates the upper and lower polariton states~\cite{TavisCummings} as $|\pm_{\bf k}\rangle=\frac{1}{\sqrt{2}}[|\mathrm{B}\rangle\otimes |0_{\bf k}\rangle\pm |G\rangle \otimes|1_{\bf k}\rangle]$ 
(which are light-matter entangled states) where $|0_{\bf k}\rangle$ and $|1_{\bf k}\rangle$ are Fock states with mode $\bf {k}$. The Rabi splitting is
\begin{equation}\label{eq:Rabi}
    \Omega_\mathrm{R}=\frac{E_{+}-E_{-}}{\hbar}=\sqrt{\frac{2\omega_{\bf k}}{\hbar \epsilon_{0}}}\sqrt{\frac{N}{\mathcal V}}{\mu_\mathrm{LL'}}\equiv 2\sqrt{N} g_\mathrm{c}\cdot\sqrt{\omega_{\bf k}},
\end{equation}
where $g_\mathrm{c} = {\mu_\mathrm{LL'}} \sqrt{1/(2\hbar\epsilon_{0}\mathcal{V})} $ is a Jaynes-Cummings~\cite{Jaynes1963PotI} type coupling strength (without the $\sqrt{\omega_{\bf k}}$-dependence). There are also $N-1$ dark states that do not mix with the photonic DOF under this simplified approximation. Details are in Supplementary Material~\cite{SI}, Sec. III. Forming the Rabi splitting/polariton states stems from a collective phenomenon, resulting in the well-known $\sqrt{N}$ dependence or $\sqrt{N/\mathcal{V}}$ dependence of $\Omega_\mathrm{R}$, which has been confirmed experimentally~\cite{Ebbesen_nanophotonics_2020}. It has been estimated that there are $N\sim 10^6-10^{12}$ molecules effectively coupled to the cavity mode~\cite{Pino2015NJP,Ribeiro2019,Horwell_2023_review} for the recent VSC experiments~\cite{Ebbesen_angew_2016,Ebbesen_nanophotonics_2020}, and $\Omega_\mathrm{R}\sim100$ cm$^{-1}$ when $\omega_{0}=1000$ cm$^{-1}$, for typical VSC experiments~\cite{Ebbesen_nanophotonics_2020,Ebbesen_angew_2019}. 
What remains largely a mystery is why the delocalized light-matter hybridization can influence chemical reaction rate constant, when a reaction is often thought of as a local phenomenon that involves breaking a bond in a molecule~\cite{Horwell_2023_review}. In addition, the entire device is kept under dark conditions such that there is no additional optical pumping~\cite{Hirai2020C, Nagarajan2021JACS, Horwell_2023_review}. As such, Ebbesen and co-workers hypothesized that the fundamental mechanism of VSC must be related to the quantum field vacuum fluctuations~\cite{Ebbesen_angew_2016, Ebbesen_science_2019}.

To provide a microscopic mechanism of VSC-modified reactions, we {\it conjecture} that the cavity modes enhance the transition from ground states to vibrationally excited state manifold of the reactant, leading to an enhancement of {\it steady-state population} of both these delocalized states on the reactant side and the excited states manifold on the product side (right well) $\{|\nu'^{j}_\mathrm{R}\rangle\}$, then relax to the vibrational ground state manifold of the product side (right well) $\{|\nu^{j}_\mathrm{R}\rangle\}$. The proposed mechanism is outlined as follows
\begin{equation}\label{scheme}
    |G\rangle \xlongrightarrow{\text{$k_1$}}\{|\nu'^{j}_\mathrm{L}\rangle\} \xlongrightarrow{\text{$k_2$}} \{|\nu'^{j}_\mathrm{R}\rangle\} \xlongrightarrow{\text{$k_3$}} \{|\nu^{j}_\mathrm{R}\rangle\}.
\end{equation}
When the molecular system is originally in the Kramers low friction regime (before the Kramers turnover~\cite{RMP_1990, Pollak1989JCP}, or so-called the energy diffusion limit), the cavity enhancement of the rate constant $k_{1}$ will occur~\cite{Sun2022JPCL, Arkajit_jpcl2022, Wang2022Collective, Wang2022JPCL, Mondal2022JCP, Arkajit_2022}. When explicitly assuming that $k_{1}\ll k_{2},k_{3}$, then $|G\rangle \xlongrightarrow{\text{$k_1$}}\{|\nu'^{j}_\mathrm{L}\rangle\}$ is the {\it rate-limiting step}, and the population of the intermediate states will reach a steady-state behavior. As such, because of the steady-state approximation, the overall rate constant is
\begin{equation}\label{rate-relation}
    k\approx k_1=k_{0}+ k_\mathrm{VSC} \ll k_{2},k_{3},
\end{equation}
where $k_{0}$ is the chemical reaction rate constant outside the cavity, and $k_\mathrm{VSC}$ accounts for the pure cavity-induced effect. Note that Eq.~\ref{rate-relation} assumes that the pure cavity effect $k_\mathrm{VSC}$ can be added with $k_{0}$, which is a {\it fundamental assumption} in the current theory. 

To quantitatively express $k_\mathrm{VSC}$, we analyze the overall effect of the cavity and the photon-loss bath environment on the molecular systems by performing a normal mode transformation~\cite{Leggett_1984,Ambegaokar_1985, Miller_2001} to the Hamiltonian in Eq.~\ref{Ham} and obtaining an effective Hamiltonian, where now the cavity modes $\{\hat{q}_{\bf k}\}$ and the photon bath $\{\hat{x}_{{\bf k},\zeta}\}$ (described by $\hat{H}_\mathrm{loss}$) are transformed into effective photonic normal mode coordinates $\{\hat{\tilde{x}}_{{\bf k},\zeta}\}$, that are collectively coupled to the system DOFs through the following term
\begin{equation}\label{int}
    \hat{H}_\mathrm{LM}= \mathcal{\hat{S}} \otimes \hat{F}_\mathrm{eff},
\end{equation}
where $\mathcal{\hat{S}} \equiv \sum_{j=1}^{N}\mu(\hat{R}_{j})\cdot\cos\varphi_{j}$ is the collective system operator, $\hat{F}_\mathrm{eff} = \sum_{{\bf k}, \zeta} \tilde{c}_{{\bf k},\zeta}\hat{\tilde{x}}_{{\bf k},\zeta}$ is the stochastic force exerted by the effective bath, $\{\hat{\tilde{x}}_{{\bf k},\zeta}\}$ are the normal modes of $\{\hat{q}_{\bf k}, \hat{x}_{{\bf k},\zeta}\}$, and the coupling constants $\tilde{c}_{{\bf k},\zeta}$ as well as bath frequencies $\tilde{\omega}_{{\bf k},\zeta}$ are characterized by an effective spectral density $J_{\mathrm{eff}}(\omega)=\sum_{\mathbf{k}} \frac{ \lambda^2_{\mathrm{c}} \omega^2_\mathbf{k}  \tau^{-1}_\mathrm{c} \omega}{\left( \omega^2_\mathbf{k} - \omega^2 \right)^2 + \tau^{-2}_\mathrm{c} \omega^2}$, where $\tau_\mathrm{c}$ is the cavity lifetime. See Supplemental Material~\cite{SI}, Sec. IV. Under the continuous $k_{\parallel}$ limit, one can replace the sum with an integral, $\sum_{\bf k}f({\bf k})\to \int dk g(k) f(k)$, where $g(k)$ is the density of states for the modes according to the dispersion relation in Eq.~\ref{eq:FP-freq}. This results in 
\begin{align}\label{Jeff-int}
    &J_{\mathrm{eff}}(\omega) = \frac{\omega_\mathrm{c}^2 \lambda^2_\mathrm{c}}{2 \tan \theta_{\mathrm{m}}} \int_{- \theta_{\mathrm{m}}}^{\theta_{\mathrm{m}}} d\theta\ \frac{\csc \theta}{\cos^4 \theta} \frac{\tau^{-1}_\mathrm{c} \omega}{\left( \omega^2_\mathbf{k} - \omega^2 \right)^2 + \tau^{-2}_\mathrm{c} \omega^2},
\end{align}
where $\omega^2_\mathbf{k}=\omega^2_\mathrm{c}(1 + \tan^2 \theta)$, where $\theta_{\mathrm{m}}$ is the maximum incident angle. By taking the limit of $\theta_\mathrm{m} \to \pi/2$ and evaluating the integral, only the contribution at $\theta = 0$ inside the integral survives, such that we arrive at the closed formalism
\begin{align}\label{Jeff-no_int}
    &J_{\mathrm{eff}}(\omega) = \lambda^2_\mathrm{c}\omega_\mathrm{c}^2 \frac{\tau^{-1}_\mathrm{c} \omega}{\left( \omega^2_\mathrm{c} - \omega^2 \right)^2 + \tau^{-2}_\mathrm{c} \omega^2},
\end{align}
with details in Supplemental Material~\cite{SI}, Sec. V.  

The rate constant change $k_\mathrm{VSC}$ in Eq.~\ref{rate-relation} originates from a purely cavity-induced effect, causing the promotion of the transition from $|G\rangle$ to the single excited states manifold $\{|\nu_{j}\rangle\}$. Note that the light-matter coupling term in Eq.~\ref{int} suggests that through the collective coupling between all molecules and all modes, the cavity operator $\hat{F}_\mathrm{eff}$ will mediate the transition. We use Fermi's golden rule (FGR) to estimate this transition rate constant. The coupling for this quantum transition is provided by $\hat{\mathcal S}$, the transition is mediated by the effective photon bath operator $\hat{F}_\mathrm{eff}$ and its spectral density $J_\mathrm{eff}(\omega)$ in Eq.~\ref{Jeff-int}. Because that $\hat{H}_\mathrm{LM}$ in Eq.~\ref{int} is a {\it non-local} operator that couples, the transition of $|G\rangle\to|\nu'^{j}_{L}\rangle$ are not independent pathways. As such, in FGR one needs to explicitly account for the interference of pathways~\cite{Cao_2022} such that one need to add them together before squaring them in the FGR.  This is the fundamental assumption of the theory and needs to be checked with future simulations and experiments. With FGR, we have
\begin{equation}\label{Theory}
    k_{\mathrm{VSC}} = \frac{1}{N}\frac{2}{\hbar} \Big|\Big(\sum_{k=1}^{N}\langle\nu_{k}|\hat{\mathcal S}|G\rangle\Big)\Big|^2\cdot J_{\mathrm{eff}}(\omega_{0})\cdot n(\omega_{0}),
\end{equation}
where $n(\omega_{0})=1/(e^{\beta\hbar\omega_{0}}-1)\approx e^{-\beta\hbar\omega_{0}}$ (when $\beta\hbar\omega_{0}\gg 1$ for  $\omega_{0}=1000$ cm$^{-1}$ and room temperature $1/\beta\approx 200$ cm$^{-1}$). Here, the $1/N$ factor accounts for the normalized rate constant per molecule, as we are considering the collective transition along $N$ molecules for the state $|G\rangle$ and $\{|\nu_{j}\rangle\}$. 

{\bf Resonant Effect.} Let us first focus on the $J_{\mathrm{eff}}(\omega_{0})$ term in $k_{\mathrm{VSC}}$ (Eq.~\ref{Theory}) expressed as
\begin{align}\label{Jeff-w0}
&J_{\mathrm{eff}}(\omega_0) = \lambda^2_\mathrm{c}\omega_\mathrm{c}^2 \frac{\tau^{-1}_\mathrm{c} \omega_0}{\left( \omega^2_\mathrm{c} - \omega^2_0 \right)^2 + \tau^{-2}_\mathrm{c} \omega^2_0}.
\end{align}
It is self-evident that the peak of this function is located at $\omega_\mathrm{c}=\omega_{0}$ for $k_{\parallel}=0$, agreeing with the experimental observation~\cite{Ebbesen_angew_2016, Hirai2020C, Nagarajan2021JACS, Simpkins-review}. Thus, the VSC-modified rate constant only occurs when $\omega_\mathrm{c}=\omega_{0}$. This is because of the Van-Hove type singularity~\cite{VanHove} in the density of states, $g({\bf k})$, which manifests itself as the $\csc\theta$ term in Eq.~\ref{Jeff-int}, such that the integral only survives and gives a finite value at $\theta=0$, and at $\theta >0$ the integral becomes vanishingly small (see Fig.~S2 of the Supplemental Material~\cite{SI}). The condition for observing the Rabi splitting, on the other hand, is $\omega^2_\mathbf{k}=\omega^2_\mathrm{c}(1 + \tan^2 \theta)=\omega_{0}$ for any $\theta\ge 0$. Although those modes with $\theta>0$ do not contribute to $k_\mathrm{VSC}$,  the mode density is finite and there will {\it always} be a mode available that satisfies $\omega^2_\mathbf{k}=\omega_{0}$, generating Rabi splitting at $\theta>0$. 

{\bf Collective Effect.} Let us focus on the square coupling matrix elements, which should explain the collective effect. If all the molecules' dipoles are perfectly aligned with the cavity TE polarization, then $\cos \varphi_{j}=1$ for all molecule $j$. This means that $\hat{\mathcal S}=\sum_{j}\mu(\hat{R}_{j})$. Evaluating Eq.~\ref{Theory} leads to 
\begin{align}\label{FGR-aligned}
    k_\mathrm{VSC}&=\frac{2}{\hbar}N\mu_{\mathrm{LL}'}^2 \cdot J_{\mathrm{eff}}(\omega_{0}) \cdot n(\omega_{0})\\
    &= {4 N g^2_\mathrm{c} \omega_\mathrm{c}^2} \cdot \frac{\tau^{-1}_\mathrm{c} \omega_{0}}{\left( \omega^2_\mathrm{c} - \omega_{0}^2 \right)^2 + \tau^{-2}_\mathrm{c} \omega_{0}^2}\cdot e^{-\beta\hbar\omega_{0}},\notag
\end{align}
where in the second line we have explicitly approximated $n(\omega_{0}) \approx e^{-\beta\hbar\omega_{0}}$. As a special case, when $\omega_\mathrm{c}=\omega_0$, Eq.~\ref{FGR-aligned} becomes $k_\mathrm{VSC}=\Omega^2_\mathrm{R}\tau_\mathrm{c}e^{-\beta\hbar\omega_{0}}$, where $\Omega_\mathrm{R}=2\sqrt{N} g_\mathrm{c}\cdot\sqrt{\omega_{0}}$. The current theory of the VSC enhanced rate constant is thus $k/k_{0}=1+k_\mathrm{VSC}/k_{0}$, with $k_\mathrm{VSC}$ expressed in Eq.~\ref{FGR-aligned}. The cavity quality factor is often defined as $Q=\tau^{-1}_\mathrm{c} \omega_{0}$ for the resonant condition. For the recent VSC experiment by Ebbesen~\cite{Ebbesen_science_2019}, the typical values for these parameters are $\tau_\mathrm{c}\approx100$ fs (reading from a width of the cavity transmission spectra of $\Gamma_\mathrm{c}\approx 53$ cm$^{-1}$). If the cavity frequency $\omega_\mathrm{c}=\omega_{0}=1200$ cm$^{-1}$, then the quality factor is $Q\approx 22.6$. 
The theory predicts the resonance effect ($\omega_\mathrm{c}=\omega_{0}$) with the requirement of $k_{\parallel}=0$ and the collective effect ($k_\mathrm{VSC}\propto N g^2_\mathrm{c}$). The mechanism is governed by thermal activation ($\propto e^{-\beta \hbar\omega_{0}}$).

Unfortunately, if we consider the fully isotropically disordered dipoles, such that under an ensemble average, the angle distribution becomes $\langle \cos\varphi_{j}\cos\varphi_{k}\rangle=\frac{1}{3}\delta_{jk}$, then the ensemble average of the coupling square matrix elements become
$\langle(\sum_{k=1}^{N}\langle\nu_{k}|\hat{\mathcal S}|G\rangle|)^2\rangle=\mu^2_{\mathrm{LL}'}\langle|\sum_{j}\cos\varphi_{j}|^2\rangle=\mu^2_{\mathrm{LL}'}\cdot N/3$, which will not give rise to any collective enhancement due to the normalization term $1/N$ in Eq.~\ref{Theory}. As such, the current theory can not predict any collective effect if considering the fully isotropic distribution of the dipole, which was assumed to be the case for most of the VSC experiments~\cite{Horwell_2023_review}. Connecting to the recent theory development of polariton mediated exciton energy transfer theory~\cite{Cao_2022}, the fully aligned dipole is equivalent to the coherent pathway picture and the fully isotropic dipoles corresponds to the incoherent picture~\cite{Cao_2022}. The current theory relies on the fully coherent energy excitation picture~\cite{Cao_2022}. In order to explain non-vanishing results for the isotropic dipoles, one either has to introduce new physics or acknowledge that $\langle \cos\varphi_{j}\cos\varphi_{k}\rangle=\frac{1}{3}\delta_{jk}$ is not the case inside the cavity~\cite{philbin_2022}.


For the subsequent steps for the reaction $\{|\nu'^{j}_\mathrm{L}\rangle\} \xlongrightarrow{\text{$k_2$}} \{|\nu'^{j}_\mathrm{R}\rangle\}$, coupling to the cavity should not cast any {\it additional} influence of $k_{2}$, due to the local nature of the chemistry. A simple argument is provided in Supplemental Material~\cite{SI}, Sec. II. For the typical VSC experiments~\cite{Ebbesen_angew_2019, Lather_2022,Ebbesen_angew_2016,Ebbesen_nanophotonics_2020,Simpkins2023}, the maximum rate change is $\sim 5$ times compared to the outside cavity case~\cite{Hirai2020C}. As long as $k_{2}>k_{1}\approx 5 k_{0}$, $k_{2}$ will not become a bottleneck that ruins the steady-state approximation. Considering that the excited state tunneling is usually fast due to the large overlap between the $|\nu'_\mathrm{L}\rangle$ and $|\nu'_\mathrm{R}\rangle$ (that gives rise to tunneling coupling), this is a reasonable assumption of the conjectured mechanism in Eq.~\ref{scheme}. Similarly, for the final step, $\{|\nu'^{j}_\mathrm{R}\rangle\} \xlongrightarrow{\text{$k_3$}} \{|\nu^{j}_\mathrm{R}\rangle\}$ there is no additional cavity modification (if the product is not coupled to the cavity).

{\bf High $\tau_\mathrm{c}$ limit.} Note that Eq.~\ref{Theory} is only valid for the low $\tau_\mathrm{c}$ limit. For $\tau_\mathrm{c}\to \infty$ limit, Eq.~\ref{Theory} will diverge due to the infinitely narrow $J_\mathrm{eff}(\omega_{0})$. The actual rate constant, however, will not diverge due to the phonon broadening (from of the $\hat{H}_{\nu}$ term in Eq.~\ref{Ham}). From analyzing the role of the phonon coupling, the rate constant should be expressed as 
\begin{equation}\label{convol}
\tilde{k}_\mathrm{VSC}=\int_{0}^{\infty} d\omega~ k_\mathrm{VSC}(\omega) G(\omega-\omega_{0}),
\end{equation}
where $k_\mathrm{VSC}(\omega) $ is expressed in Eq.~\ref{FGR-aligned}, $G$ is a Gaussian broadening function, with the width controlled by the phonon broadening $\sigma^2 = \epsilon^2_{z}\cdot \frac{1}{\pi} \int_{0}^\infty d\omega~ J_\nu (\omega) \coth(\beta \omega / 2)$, where $\epsilon_{z} = \langle \nu'_\mathrm{L}|\hat{R}| \nu'_\mathrm{L}\rangle-\langle\nu_\mathrm{L}|\hat{R}|\nu_\mathrm{L}\rangle$. One can further show that under the limit that $\tau^{-1}_\mathrm{c}\gg \sigma$, Eq.~\ref{convol} reduces to Eq.~\ref{Theory}. Under the limit that $\tau^{-1}_\mathrm{c}\ll \sigma$, Eq.~\ref{FGR-aligned} reduces to  $\tilde{k}_\mathrm{VSC} \approx2\pi{N g^2_\mathrm{c} \omega_\mathrm{c}} G(\omega_\mathrm{c}-\omega_{0}) e^{-\beta \hbar \omega_0}$, where the Gaussian function gives the resonant effect. Details of derivation, as well as the numerical behavior of Eq.~\ref{convol} (Fig. S3), are provided in Supplemental Material~\cite{SI}, Sec. VI.

{\bf Reaction Rate.} We want to point out that the reaction rate $\mathcal {R}=(k_{0}+k_\mathrm{VSC})\cdot N$ is expressed as  
\begin{equation}\label{rate}
\mathcal{R}=Nk_{0}+{4 N^2 g^2_\mathrm{c} \omega_\mathrm{c}^2} \cdot \frac{\tau^{-1}_\mathrm{c} \omega_{0}}{\left( \omega^2_\mathrm{c} - \omega_{0}^2 \right)^2 + \tau^{-2}_\mathrm{c} \omega_{0}^2}\cdot e^{-\beta\hbar\omega_{0}},
\end{equation}
which predicts a fundamental change in the scaling with respect to the number of molecules (concentration). Note that not to be confused between the content of rate $\mathrm{R}$ and rate constant $k$ (per molecule rate). Outside the cavity, $k_\mathrm{VSC}=0$, such that $\mathcal{R}=(k_{0}+k_\mathrm{VSC})\cdot N$ for the simple unimolecular reaction we considered here. By coupling to the optical cavity, if $k_\mathrm{VSC}\ll k_{0}$, the reaction rate scales as $\mathcal{R}\propto N^2$, which is fundamentally different than outside the cavity case. In the recent VSC-enhanced experiments, one does observe that (for example, Fig.~4b in Ref.~\cite{Lather2020JPCL}) $\mathcal{R}$ changes the scaling with respect to $N$ from a liner dependence (outside the cavity) to a non-linear dependence of $N$.

\begin{figure}[htbp]
\centering
\includegraphics[width=1.0\linewidth]{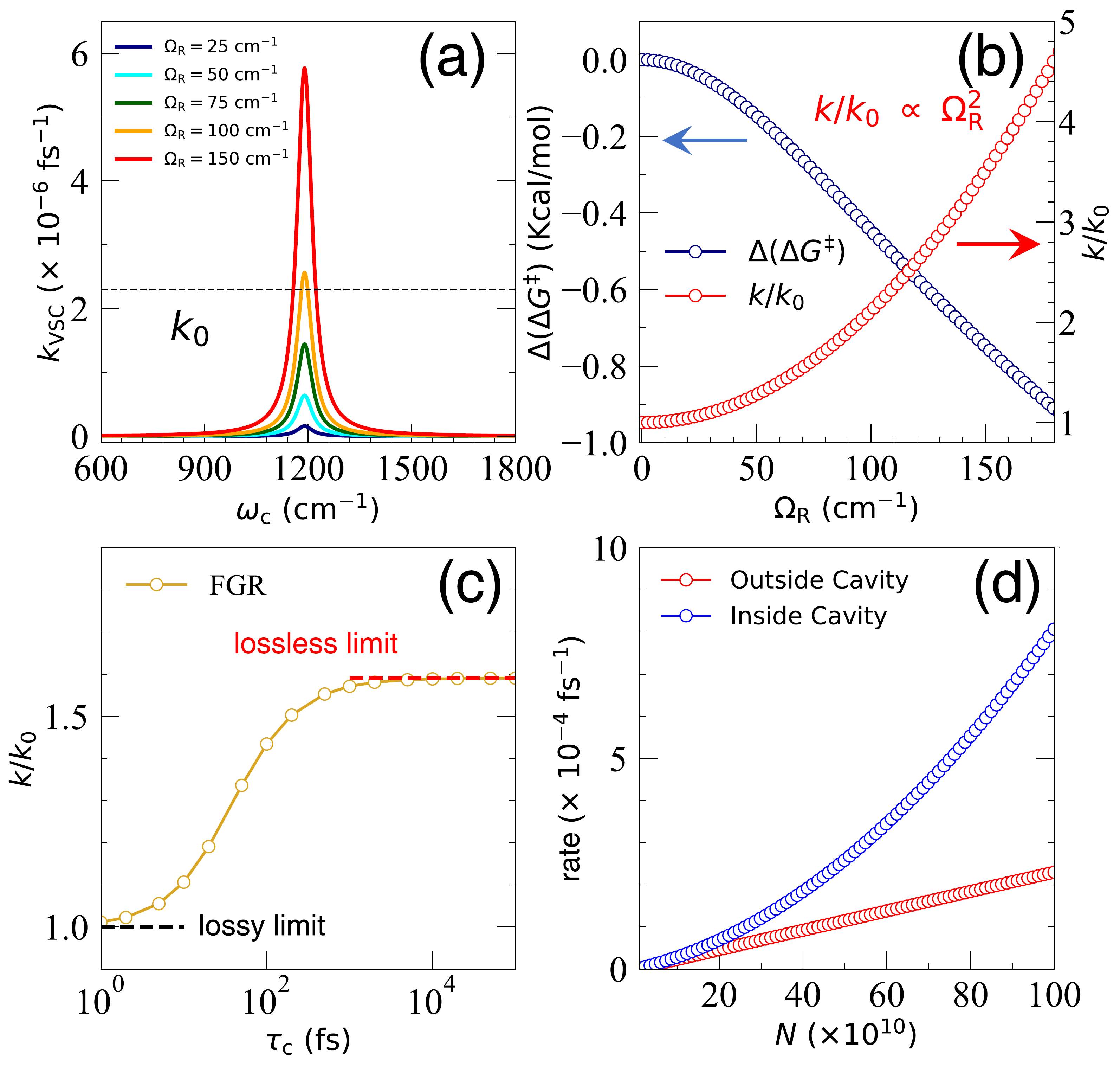}
\caption{Numerical evaluation of the $k_\mathrm{VSC}$ in Eq.~\ref{FGR-aligned}. (a) $k_\mathrm{VSC}$ as a function of $\omega_\mathrm{c}$ with different  $\Omega_\mathrm{R}$ (Eq.~\ref{eq:Rabi}). The cavity lifetime is $\tau_\mathrm{c} = 100$ fs. (b) $k/k_{0}$ (red) when $\omega_\mathrm{c}=\omega_{0}$ as a function of $\Omega_\mathrm{R}$, where $k_{0}=2.3 \times 10^{-6}$ fs$^{-1}$ was taken from the exact simulation for the same model~\cite{Arkajit_2022}. The effective change of the free energy barrier $\Delta(\Delta G^{\ddagger})$ (blue) is also plotted. (c) The $k/k_{0}=1+\tilde{k}_\mathrm{VSC}/k_{0}$ using Eq.~\ref{convol}, as a function of $\tau_\mathrm{c}$. (d) Reaction Rate $\mathcal{R}$ as a function of $N$ for outside the cavity (red) and inside the cavity case (blue).}
    \label{fig_theory}
\end{figure}

Fig.~\ref{fig_theory} presents the numerical results of using Eq.~\ref{FGR-aligned} (or Eq.~\ref{convol} to estimate VSC-enhanced experiments. Here, we consider the same symmetric double well model used in the recent work~\cite{Arkajit_2022}. Fig.~\ref{fig_theory}a present the $k_\mathrm{VSC}$ by changing $\omega_\mathrm{c}$ using Eq.~\ref{FGR-aligned}. The result obtained from Eq.~\ref{convol} is nearly identical to Eq.~\ref{FGR-aligned} due to the low cavity lifetime $\tau_\mathrm{c}=100$ fs. Fig.~\ref{fig_theory}a presents the sharp resonance behavior of the VSC-modified rate constant profile at $\omega_\mathrm{c}=\omega_{0}=1190$ cm$^{-1}$. Fig.~\ref{fig_theory}b highlights the quadratic dependence of $k/k_{0}\propto \Omega^2_\mathrm{R}$, which manifests into the effective barrier height $\Delta(\Delta G^\ddagger)\equiv-k_\mathrm{B}T\ln \left({k}/{k_0}\right)$ if one chooses to interpret the VSC effect from transition state theory~\cite{Ebbesen_nanophotonics_2020, Ebbesen_angew_2016, Ebbesen_angew_2019} (which is inappropriate). It is clear that an $\Omega_\mathrm{R}=150$ cm$^{-1}$ can cause a $\Delta(\Delta G^\ddagger)\approx$ 0.8 Kcal/mol (or 3.3 KJ/mol), explaining the similar basic experimental trend~\cite{Hirai2020C}. It is also clear why $\Delta(\Delta G^\ddagger)$ has to be non-linearly related to $\Omega_\mathrm{R}$. Fig.~\ref{fig_theory}c presents the $k_\mathrm{VSC}$ when $\omega_\mathrm{c}=\omega_{0}$ as a function of cavity lifetime $\tau_\mathrm{c}$ (on a log scale in $\tau_\mathrm{c}$). The rate of constant modification changes in a sigmoid trend, which is subject to future experimental verification. 
Fig.~\ref{fig_theory}d presents the reaction rate $\mathcal{R}$ as a function of $N$ (using Eq.~\ref{rate}), under a coupling strength $g_\mathrm{c}$ for $\Omega_\mathrm{R}=150$ cm$^{-1}$ when $N=10^{12}$. As one expected, outside the cavity, $R=Nk_{0}$ and scales linearly with $N$. Inside the cavity, when the rate constant is dominated by $k_\mathrm{VSC}$, $\mathcal{R}$ starts to show $N^2$ scaling. Future experiments are needed to confirm this non-linear trend.

To summarize, we conjecture that the fundamental mechanism of the VSC-influenced rate process is related to the $|G\rangle \xlongrightarrow{\text{$k_1$}}\{|\nu'^{j}_\mathrm{L}\rangle\}$ process, which is the rate-limiting step. The pathways are constructively interfered with each other due to the non-local light-matter coupling operator (Eq.~\ref{int}). That said, the subsequent steps do encounter the local nature of the chemical reaction and will not be influenced by coupling to the cavity. So long as $k_{1}$ is rate limiting, the many-body, coherent cavity modification of vibrational excitation associated with $k_1$ will {\it manifest in the local chemical reaction} and influence the apparent rate constant. This is our mechanistic explanation of how the delocalized nature of light-matter interaction influences local bond-breaking chemistry. Any reaction that does not satisfies the condition  $k_{1}\ll k_{2},k_{3}$  will likely give negative results of VSC modifications. We hope the current theory could  offer valuable insights into the fundamental mechanism of vibrational Polariton Chemistry.

\begin{acknowledgments}
{\bf Acknowledgement}. This work was supported by the National Science Foundation Award under Grant No. CHE-2244683. P.H. appreciates the support of the Cottrell Scholar Award (a program by the Research Corporation for Science Advancement).  M.A.D.T. appreciates the support from the National
Science Foundation Graduate Research Fellowship Program under Grant No. DGE-1939268. We appreciate valuable discussions with Jino George, Tao Li, Eric Koessler, and Arkajit Mandal.
\end{acknowledgments}


\providecommand{\noopsort}[1]{}\providecommand{\singleletter}[1]{#1}%

\end{document}


\title[]{Supplementary Material for \\
Microscopic Theory of Vibrational Polariton Chemistry}

\author{Wenxiang Ying}
\affiliation{Department of Chemistry, University of Rochester, Rochester, NY 14627, USA}
\author{Michael A.D. Taylor}
\affiliation{The Institute of Optics, Hajim School of Engineering, University of Rochester, Rochester, NY 14627, USA}
\author{Pengfei Huo}
\email{pengfei.huo@rochester.edu}
\affiliation{Department of Chemistry, University of Rochester, Rochester, NY 14627, USA}
\affiliation{The Institute of Optics, Hajim School of Engineering, University of Rochester, Rochester, NY 14627, USA}

\maketitle

\section{I. Details of the VSC Hamiltonian}
We start with the Pauli-Fierz Hamiltonian of many molecules coupled to many modes under the dipole approximation. This Hamiltonian is obtained by performing the Power-Zienau-Woolley (PZW) gauge transformation~\cite{Power1959PTRSA, CohenTannoudji1997, Woolley1974JPBAMP} on the minimum coupling Hamiltonian. The details can be found in Ref.~\citenum{Mandal2022C} (Sec. 2.6) or Ref.~\citenum{Keeling2012} (Chapter 2.2). The Hamiltonian is then further projected on the ground electronic states of all molecules.

The total Hamiltonian is expressed as
\begin{align} \label{eq:GenPF}
&\hat{H}=\hat{H}_\mathrm{M} + \hat{H}_\mathrm{ph}+\hat{H}_\mathrm{loss}\\ 
&+ \sum_\mathbf{k} \Big[ \sqrt{\frac{\hbar \omega_{\mathbf{k}}}{2}} \lambda_\mathrm{c} \sum_j (\hat{a}^{\dagger}_{\bf k}e^{- i {\mathbf{k}} \cdot \bar{\boldsymbol x}_j} + \hat{a}_{\bf k} e^{i {\mathbf{k}} \cdot \bar{\boldsymbol x}_j}) (\hat{\text e}_{\mathbf{k}} \cdot \hat{\boldsymbol \mu}_j (\hat{\bf R}_j)) + \frac{\lambda_\mathrm{c}^2}{2} \sum_{i,j} (\hat{\text e}_{\mathbf{k}} \cdot \hat{\bm{\mu}}_{i}(\hat{\bf R}_i)) (\hat{\text e}_{\mathbf{k}} \cdot \hat{\bm{\mu}}_{j}(\hat{\bf R}_j)) e^{-i \mathbf{k} \cdot (\bar{\boldsymbol x}_i - \bar{\boldsymbol x}_j)} \Big],\notag
\end{align}
where the $\{i,j\}$ iterates over the molecules in the cavity, $\bar{\boldsymbol x}_j$ is the center of mass of the $j_\mathrm{th}$ molecule, $\hat{H}_\mathrm{M}$ is the bare matter Hamiltonian, $\hat{H}_\mathrm{ph}$ is the pure photonic Hamiltonian. Further, $\hat{H}_\mathrm{M}$ is the molecular Hamiltonian 
\begin{equation}
\hat{H}_\mathrm{M}=\sum_{j=1}^{N}\Big(\frac{\hat{P}_{j}^2}{2M} + V(\hat{R}_{j})\Big)+\hat{H}_{\nu},
\end{equation}
where $\hat{R}_{j}$ is the reaction coordinate for the $j_\mathrm{th}$ molecule, $V(\hat{R})$ is the ground state potential for all reaction molecules (typically double well potential), and $\mu(\hat{R}_{j})$ is the dipole associated with the ground electronic state (electronic permanent dipole). In this work, we have explicitly ignored the interactions among molecules and treated them as independent, identical molecules. We further introduced the effects of a phonon bath to the Hamiltonian with the system-bath term, which reads as
\begin{align}\label{Hnu}
\hat{H}_\nu = \frac{1}{2} \sum_{j}\sum_{\zeta} \Big[ \hat{p}^2_{j, \zeta} + \omega^2_{j, \zeta} \Big(\hat{x}_{j, \zeta} - \frac{c_{j, \zeta}}{\omega^2_{j, \zeta}} \hat{R}_j \Big)^2 \Big],
\end{align}
where $\{ \hat{x}_{j, \zeta}$, $\hat{p}_{j, \zeta} \}$ are the mass-weighted coordinate and momentum pair of the $\{j, \zeta\}$-th bath oscillator that directly couples to the reaction coordinate of molecule $j$. The $j$-th phonon bath as well as its coupling to the reaction coordinate $R_j$ can be described by the spectral density function~\cite{Caldeira_Leggett_1983}
\begin{align}
    J_\nu (\omega) = \frac{\pi}{2} \sum_j \sum_\zeta \frac{c^2_{j, \zeta}}{\omega_{j, \zeta}} \delta(\omega - \omega_{j, \zeta}),
\end{align}
where $\omega_{j, \zeta}$, $c_{j, \zeta}$ are the oscillator frequencies and coupling coefficients, respectively. Note that we have assumed identical spectral density for all molecule $j\in[1,N]$. 

By introducing the photon mode coordinate and momentum operators 
\begin{align}
    \hat{q}_{\bf k} = \sqrt{\hbar/2\omega_{\bf k}}(\hat{a}_{\bf k}^{\dagger} + \hat{a}_{\bf k}),~~~~~~ \hat{p}_{\bf k} = i\sqrt{\hbar\omega_{\bf k}/2}(\hat{a}_{\bf k}^{\dagger} - \hat{a}_{\bf k}),
\end{align}
or inversely, the field operators
\begin{align}
    \hat{a}_{\bf k} = \sqrt{\frac{\omega_\mathbf{k}}{2\hbar}} \hat{q}_{\bf k} + i \sqrt{\frac{1}{2\hbar \omega_\mathbf{k}}} \hat{p}_{\bf k},~~~~~~ \hat{a}_{\bf k}^{\dagger} = \sqrt{\frac{\omega_\mathbf{k}}{2\hbar}} \hat{q}_{\bf k} - i \sqrt{\frac{1}{2\hbar \omega_\mathbf{k}}} \hat{p}_{\bf k},
\end{align}
Eq.~\ref{eq:GenPF} can be alternatively expressed as
\begin{align} \label{GenPF_concise}
    \hat{H} = \hat{H}_\mathrm{M} + \frac{1}{2} \sum_\mathbf{k} \left[(\hat{p}_{\bf k} - \lambda_\mathrm{c} \hat{\Pi}_\mathbf{k})^2 + (\omega_{\bf k} \hat{q}_{\bf k} + \lambda_\mathrm{c} \hat{\mathcal{S}}_\mathbf{k})^2 \right]+\hat{H}_\mathrm{loss},
\end{align}
where the collective system operators are defined as
\begin{subequations}
\begin{align}
    \hat{\Pi}_\mathbf{k} &= \sum_j (\hat{\text e}_{\mathbf{k}} \cdot \hat{\boldsymbol \mu}_j (\hat{\bf R}_j)) \sin(\mathbf{k} \cdot \bar{\boldsymbol x}_j), \\
    \hat{\mathcal{S}}_\mathbf{k} &= \sum_j (\hat{\text e}_{\mathbf{k}} \cdot \hat{\boldsymbol \mu}_j (\hat{\bf R}_j)) \cos(\mathbf{k} \cdot \bar{\boldsymbol x}_j).
\end{align}
\end{subequations}


To account for cavity loss, we further introduce the loss Hamiltonian which is also based on the system-bath model, defined as
\begin{align}
    \hat{H}_\mathrm{loss} = \frac{1}{2} \sum_{{\bf k}, \zeta} \Big[ \hat{p}^2_{{\bf k}, \zeta} + \omega^2_{{\bf k}, \zeta} \Big(\hat{x}_{{\bf k}, \zeta} - \frac{c_{{\bf k}, \zeta}}{\omega^2_{{\bf k}, \zeta}} \hat{q}_{\bf k} \Big)^2 \Big],
\end{align}
where $\{\hat{x}_{{\bf k},\zeta}, \hat{p}_{{\bf k}, \zeta}\}$ are the mass-weighted coordinate and momentum operators of the $\{ {\bf k}, \zeta \}$-th non-cavity bath mode, respectively, which directly couple to the photon mode coordinate operator $\hat{q}_\mathbf{k}$. The 
$\bf k$-th loss bath as well as its coupling to the photon mode coordinate operator $\hat{q}_\mathbf{k}$ are described by the spectral density function
\begin{align} \label{spectral_density_loss}
    J_\mathrm{loss}(\omega, {\bf k}) = \frac{\pi}{2} \sum_\zeta \frac{c^2_{{\bf k}, \zeta}}{\omega_{{\bf k}, \zeta}} \delta(\omega - \omega_{{\bf k}, \zeta}),
\end{align}
where $\omega_{{\bf k}, \zeta}$, $c_{{\bf k}, \zeta}$ are the oscillator frequencies and coupling coefficients, respectively.

Based on the above discussions, the collective PF Hamiltonian in Eq.~\ref{GenPF_concise} is expressed in the projected subspace as
\begin{align} \label{GenPF_gauge_form}
    \hat{H} = \sum_j \frac{\hat{P}_j}{2M_j} + E_g(R_j) + \hat{H}_\nu + \frac{1}{2} \sum_\mathbf{k} \left[(\hat{p}_{\bf k} - \lambda_\mathrm{c} \hat{\Pi}_\mathbf{k})^2 + (\omega_{\bf k} \hat{q}_{\bf k} + \lambda_\mathrm{c} \hat{\mathcal{S}}_\mathbf{k})^2 \right] + \hat{H}_\mathrm{loss},
\end{align}
where
\begin{subequations}
\begin{align}
    \hat{\Pi}_\mathbf{k} &= \sum_j \mu_j (\hat{R}_j) \cdot \cos \varphi_j \cdot \sin(\mathbf{k} \cdot \bar{\boldsymbol x}_j), \\
    \hat{\mathcal{S}}_\mathbf{k} &= \sum_j \mu_j (\hat{R}_j) \cdot \cos \varphi_j \cdot \cos(\mathbf{k} \cdot \bar{\boldsymbol x}_j), \label{eq:SK}
\end{align}
\end{subequations}
where $\mu_j (\hat{R}_j)$ is the ground state permanent dipole moment of molecule $j$, and $\varphi_j$ is the relative angle between the dipole vector and the field polarization direction $\hat{\text e}_{\mathbf{k}}$. 

For simplicity, in this work we assume the {\it long wavelength approximation}, the transverse fields can be treated as spatially uniform, {\it i.e.}, $e^{i {\bf k}\cdot{\bf r}}\approx1$, such that 
\begin{equation}\label{eq:lwa}
    \hat{\bf A}_{\perp}({\bf r})\approx \hat{\bf A}_{\perp} = \sum_{\bf k} \frac{\hat{{\mathcal{\text e}}}_{\bf k}}{\omega_{\bf k}} \sqrt{\frac{\hbar\omega_{\bf k}}{2\varepsilon_{0} \mathcal{V}}} (\hat{a}_{\bf k}  +  \hat{a}^{\dagger}_{\bf k}),
\end{equation}
which leads to $\cos(\mathbf{k} \cdot \bar{\boldsymbol x}_j) = 1$, $\sin(\mathbf{k} \cdot \bar{\boldsymbol x}_j)=0$, $\hat{\Pi}_\mathbf{k}=0$ and $\hat{\mathcal{S}}_\mathbf{k} = \sum_j \mu_j (\hat{R}_j) \cdot \cos \varphi_j$. Then the collective PF Hamiltonian of Eq.~\ref{GenPF_gauge_form} can be further simplified as 
\begin{align} \label{GenPF_LWA}
    \hat{H}_\mathrm{PF}^{[N]} = \sum_j \frac{\hat{P}_j}{2M} + V(R_j) + \hat{H}_\nu + \frac{1}{2} \sum_\mathbf{k} \Big[\hat{p}_{\bf k}^2 + \omega_{\bf k}^2 \Big(\hat{q}_{\bf k} +  \frac{\lambda_\mathrm{c}}{ \omega_{\bf k}} \sum_j \mu (\hat{R}_j) \cdot \cos \varphi_j \Big)^2 \Big] + \hat{H}_\mathrm{loss},
\end{align}
which is the VSC Hamiltonian in Eq.~3 of the main text. To summarize, we have explicitly assumed that there are no inter-molecular interactions, as well as the long wavelength approximation. Their presence might influence the final rate constant expression $k_\mathrm{VSC}$, however, they are not playing any dominant role in order to provide a theory as we have shown here to capture resonant effect and collective effect.

\section{II. Details of the Molecular Model}\label{model}
To model how VSC influences chemical reactions, we are particularly interested in the one-dimensional double-well (DW) potential~\cite{Makri_1994, QShi_2011}
\begin{align}\label{DBW}
    V (\hat{R}) = - \frac{M \omega^2_\mathrm{b}}{2} \hat{R}^2 + \frac{M^2 \omega^4_\mathrm{b}}{16 E_\mathrm{b}} \hat{R}^4, 
\end{align}
where $M$ is the effective mass of the reaction coordinate, $\omega_\mathrm{b}$ is the barrier frequency, and $E_\mathrm{b}$ is barrier height of the DW potential. Note that Eq.~\ref{DBW} assumes a symmetric DW potential. 

For the system (reaction coordinate), the corresponding eigenvectors $|\nu_i\rangle$ and eigenenergies $E_{i}$ are obtained by numerically solving 
\begin{equation}\label{TISE}
    \Big(\frac{\hat{P}^2}{2M} + V(\hat{R})\Big)|\nu_i\rangle= E_{i}|\nu_i\rangle,
\end{equation}
where $V(\hat{R})$ is expressed in Eq.~\ref{DBW}. These vibrational eigenstates are obtained by using discrete variable representation (DVR) basis~\cite{Miller_1992}. We \textit{diabatize} the two lowest eigenstates as
\begin{align}\label{nuLR}
    |\nu_\mathrm{L}\rangle = \frac{1}{\sqrt{2}}\big(|\nu_0\rangle + |\nu_1\rangle\big),~~~~~ |\nu_\mathrm{R}\rangle = \frac{1}{\sqrt{2}}\big(|\nu_0\rangle - |\nu_1\rangle\big),
\end{align}
which leads to two energetically degenerate diabatic states, denoted as $|\nu_\mathrm{L}\rangle$ and $|\nu_\mathrm{R}\rangle$ for state localized in the left well and state localized in the right well, respectively, both with the degenerate energy ${E}_\mathrm{L} = (E_1 + E_0) / 2$ and a small tunneling splitting $V^{0}_\mathrm{LR} = (E_1 - E_0) / 2$ (where the energy difference between $E_1$ and $E_0$ is $2V^{0}_\mathrm{LR}$). Similarly, for $\{|\nu_2\rangle, |\nu_3\rangle \}$, one can diabatize them and obtain the first excited {\it diabatic vibrational state} in the left well and right well as follows
\begin{equation}\label{nuLR-ext}
    |\nu'_\mathrm{L}\rangle = \frac{1}{\sqrt{2}}\big(|\nu_2\rangle + |\nu_3\rangle\big),~~~~~ |\nu'_\mathrm{R}\rangle = \frac{1}{\sqrt{2}}\big(|\nu_2\rangle -|\nu_3\rangle\big),
\end{equation}
with the degenerate diabatic energy ${E}_{\mathrm{L}'} = (E_3 + E_2) / 2$ and the tunneling splitting $V_\mathrm{LR} = (E_3 - E_2) / 2$. Based on the two diabatic states $|\nu_\mathrm{L}\rangle$ and $|\nu'_\mathrm{L}\rangle$ in the left well, we define the quantum vibration frequency of the reactant as 
\begin{equation}\label{omega0}
    \hbar\omega_0 \equiv {E}_{\mathrm{L}'} - {E}_\mathrm{L},
\end{equation}
which is directly related to the quantum transition of $|\nu_\mathrm{L}\rangle \to |\nu'_\mathrm{L}\rangle$. Note that the spectroscopy measurement (IR or transmission spectra) is also directly related to this frequency. 

For a practical calculation, truncation has to be made upon the number of matter states, restricting the dynamics in a relatively low energy subspace while ensuring numerical accuracy. As such, the Hamiltonian and the reaction coordinate have their matrix representations in a truncated Hilbert space. Similarly, we also have the vibrational permanent dipole associated with $|\nu_\mathrm{L}\rangle$ as $\mu_{\mathrm{L}\mathrm{L}}=\langle \nu_\mathrm{L}|\mu(\hat{R})|\nu_\mathrm{L}\rangle$, as well as for vibrationally excited states $|\nu'_\mathrm{L}\rangle$ as $\mu_{\mathrm{L}'\mathrm{L}'}=\langle \nu'_\mathrm{L}|\mu(\hat{R})|\nu'_\mathrm{L}\rangle$. These permanent dipoles might be important for computing polariton spectra under ultra-strong coupling regimes. For rate constants, we find that they might be important to provide constant shifts of vibrational states under very large coupling limits. We have ignored them for the simplicity of the theory. 

\begin{figure*}
    \centering
    \includegraphics[width=0.65\linewidth]{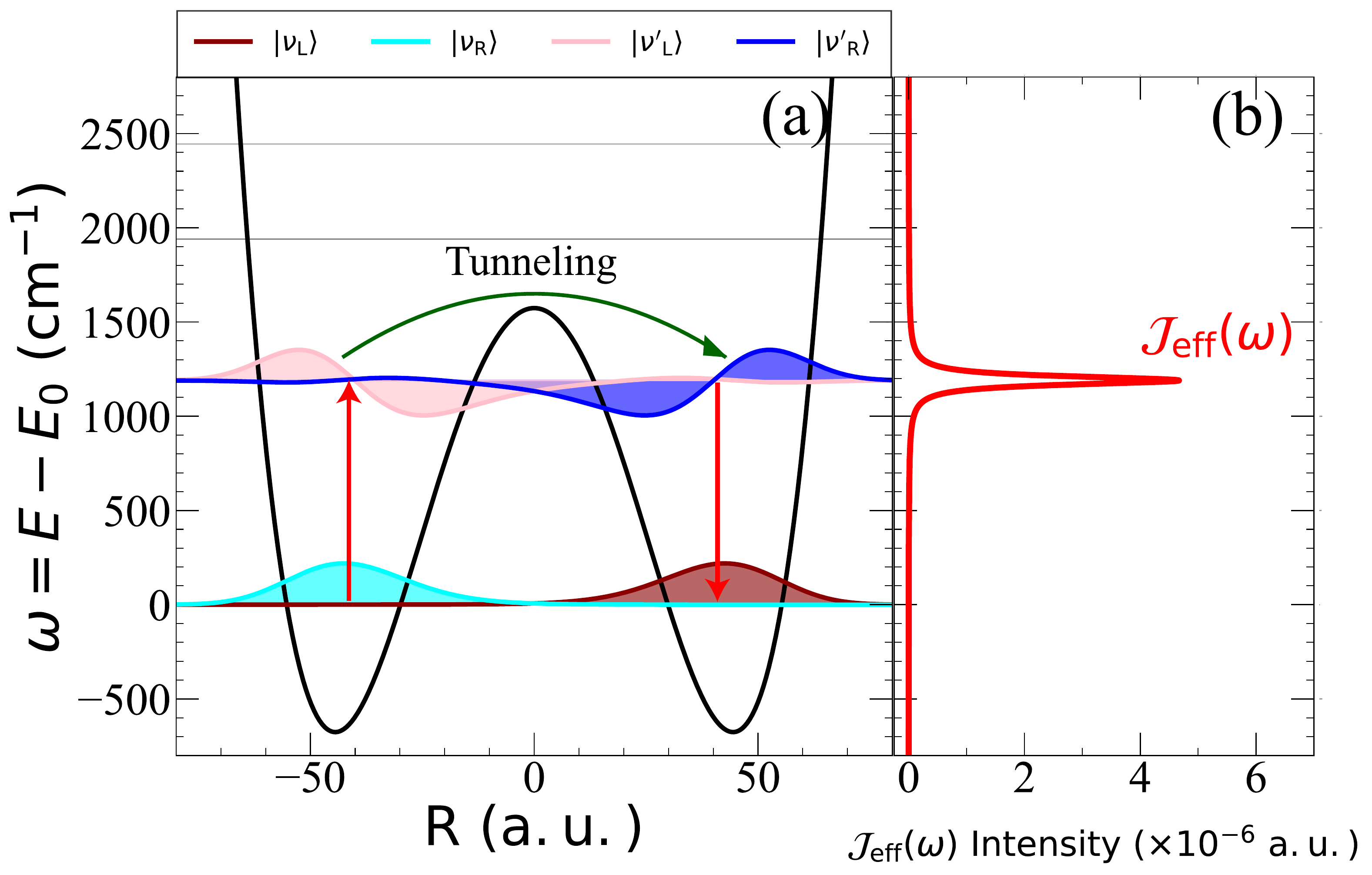}
    \caption{Schematic illustration of the ground state chemical reaction model and the environmental spectral density functions. (a) Potential energy surface for the DW model used in this work, with the plot of the first few diabatic states. The red arrows account for cavity modification effects. The ground state populations of the left well $|\nu_\mathrm{L} \rangle$ are pumped up to $|\nu'_\mathrm{L}\rangle$ states, then transfer to $|\nu'_\mathrm{R}\rangle$ via the tunneling splitting, and de-excite to the right well $|\nu_\mathrm{R} \rangle$. (b) Plot of the effective spectral density $J_{\mathrm{eff}}(\omega)$ (red curve), which corresponds to the cavity and its associated loss (see Sec.~IV for details). The parameters are taken as: the light-matter coupling strength $\lambda_\mathrm{c} = 100\ \mathrm{cm}^{-1}$, the cavity frequency $\hbar \omega_\mathrm{c} = 1190\ \mathrm{cm}^{-1}$ (in resonance), and the cavity lifetime $\tau_\mathrm{c} = 100\ \mathrm{fs}$.}
    \label{fig_mechanism}
\end{figure*}

Fig.~\ref{fig_mechanism} provides a schematic illustration of the ground state chemical reaction model (single molecule) and the environmental spectral density functions outside and inside the cavity, respectively. Here, we use the parameters $E_\mathrm{b} = 2250\ \mathrm{cm}^{-1}$, and $\hbar \omega_\mathrm{b} = 1000\ \mathrm{cm}^{-1}$.~\cite{Arkajit_2022} The eigenstates are obtained with the sinc-DVR basis with 1001 grid points in the range of $- 100 \leq R \leq 100$, then diabatized according to Eq.~\ref{nuLR} and \ref{nuLR-ext}. Note that because $|\nu'_\mathrm{L}\rangle$ and $|\nu'_\mathrm{R}\rangle$ are very close to the top of the barrier, they are not as well localized as $|\nu_\mathrm{L}\rangle$ and $|\nu_\mathrm{R}\rangle$. 

Fig.~\ref{fig_mechanism}a presents the first few vibrational states of the DW model, denoted as $|\nu_\mathrm{L}\rangle$, $|\nu_\mathrm{R}\rangle$, $|\nu'_\mathrm{L}\rangle$, $|\nu'_\mathrm{R}\rangle$. The red arrows indicate the potential effect of the cavity modifying vibrational state transitions, and the green arrow right above the barrier denotes to the fast dissipative tunneling process from $|\nu'_\mathrm{L}\rangle$ to $|\nu'_\mathrm{R}\rangle$. 

Fig.~\ref{fig_mechanism}b shows the effective spectral density $J_{\mathrm{eff}}(\omega)$ (red) which resembles the Brownian oscillator spectral density that roughly centers at $\omega_\mathrm{c}$ when the loss bath is Markovian. It is found that $J_{\mathrm{eff}}(\omega)$ can accelerate the state-to-state quantum transitions $|\nu_\mathrm{L}\rangle\to|\nu'_\mathrm{L} \rangle$ and $|\nu_\mathrm{R}\rangle\to|\nu'_\mathrm{R} \rangle$ (as is indicated by the red arrows in Fig.~\ref{fig_mechanism}a) when its peak frequency is in resonance with the quantum vibrational frequency $\omega_{0}$, bringing about resonance enhancement effects. Details about the effective spectral density can be found later in Sec.~IV. To be more clear, we briefly summarize the major parameters for the model in Table~\ref{tab_par}.~\cite{Arkajit_2022}
\begin{table}[htbp]
    \caption{Table of major parameters}
    \begin{tabular*}{1.0\columnwidth}{c @{\extracolsep{\fill}} c @{\extracolsep{\fill}} c @{\extracolsep{\fill}} c @{\extracolsep{\fill}} c}
        \hline\hline
        Parameters of system DOFs & Notation & Value \\
        \hline
        Effective mass of the reaction coordinate & $M$ & $1\ \mathrm{a.u.}$ \\
        Barrier height & $E_{\mathrm{b}}$ & $2250\ \mathrm{cm}^{-1}$ \\
        Barrier frequency & $\hbar \omega_{\mathrm{b}}$ & $1000\ \mathrm{cm}^{-1}$\\ 
        vibration frequency (or resonance frequency) & $\hbar \omega_0$ & $1190\ \mathrm{cm}^{-1}$\\
        tunneling splitting between $|\nu_\mathrm{L}\rangle$ and $|\nu_\mathrm{R}\rangle$ & $V^0_\mathrm{LR}$ & $1.03\ \mathrm{cm}^{-1}$ \\
        tunneling splitting between $|\nu'_\mathrm{L}\rangle$ and $|\nu'_\mathrm{R}\rangle$ & $V_\mathrm{LR}$ & $47.68\ \mathrm{cm}^{-1}$ \\
        [0.5ex]
        \hline\hline
    \end{tabular*}
    \label{tab_par}
\end{table}

We further show that the subsequent step $\{|\nu'^{j}_\mathrm{L}\rangle\} \xlongrightarrow{\text{$k_2$}} \{|\nu'^{j}_\mathrm{R}\rangle\}$ will occur with the same rate constant $k_{2}$ as for the cavity free case, such that there is {\it no additional} change of this step due to coupling to the cavity. Note that outside the cavity, this rate is controlled by the tunneling-splitting coupling between $|\nu'^{j}_\mathrm{L}\rangle$ and $|\nu'^{j}_\mathrm{R}\rangle$, denoted as $V_\mathrm{LR}=\langle \nu'^{j}_\mathrm{R}|\hat{V}_{j}|\nu'^{j}_\mathrm{L}\rangle$, which is assumed to be identical for all molecules $j$. The localness of chemical reactions ensures that $\langle\nu'^{i}_\mathrm{R}|\hat{V}_{j}|\nu'^{k}_\mathrm{L}\rangle=V_\mathrm{LR}\delta_{ij}\delta_{jk}$, {\it i.e.}, the reaction occurs locally. Using similar FGR argument for the $\{|\nu'^{j}_\mathrm{L}\rangle\} \xlongrightarrow{\text{$k_2$}} \{|\nu'^{j}_\mathrm{R}\rangle\}$ transition,

We assume that by reaching the steady state populations, all $|\nu'^{j}_\mathrm{L}\rangle$ are equally populated, such that each channel should be weighted by $1/N$. Using FGR, and considering a resonant energy tunneling transiton between $|\nu'^{j}_\mathrm{L}\rangle$ and $|\nu'^{j}_\mathrm{R}\rangle$, the tunneling rate constant $k_{2}$ is
\begin{equation}\label{k2}
k_{2}\propto \frac{1}{N}\sum_{j=1}^{N} |\langle\nu'^{j}_\mathrm{R}|\hat{V}_{j}|\nu'^{j}_\mathrm{L}\rangle|^2 \propto V^2_\mathrm{RL},
\end{equation}
indicating that $k_{2}$ is identical to the single molecule tunneling rate outside the cavity. As such, the molecule-cavity interaction will not change the local bond breaking process.

\section{III. Rabi Splitting}
To quickly review the well-known results of Rabi splitting through collective light-matter couplings, let us assume all dipoles are fully aligned, such that $\cos\varphi_{j}=1$ for all $j$. We further introduce the $\sigma_{j}=|G\rangle\langle \nu_{j}|$ and $\sigma^{\dagger}_{j}=|\nu_{i}\rangle\langle G|$ as the raising and lowering operators of the molecular vibrational excitation on molecule $j$. In the single excited subspace $\mathcal{P}=|G\rangle\langle G|+\sum_{j}|\nu_{j}\rangle\langle \nu_{j}|$, the $\mu(\hat{R}_{j})$ operator becomes
\begin{equation}
    \hat{\mathcal P}\mu(\hat{R}_{j})\hat{\mathcal P}=\mu_\mathrm{LL'} \cdot (\sigma^{\dagger}_{j}+\sigma_{j}),
\end{equation}
where $\mu_\mathrm{LL'}=\langle \nu'^{j}_\mathrm{L}|\mu(\hat{R}_{j})|\nu^{j}_\mathrm{L}\rangle$ is identical to all molecules. Here, we explicitly ignored the permanent dipole contribution $\langle \nu^{j}_\mathrm{L}|\hat{R}_{j}|\nu^{j}_\mathrm{L}\rangle$ and $\langle \nu'^{j}_\mathrm{L}|\hat{R}_{j}|\nu'^{j}_\mathrm{L}\rangle$. They could be important for computing the polariton eigenspectrum when the coupling strength $\lambda_\mathrm{c}$ is very large.

Using the above notations, one can rewrite the light-matter coupling term as
\begin{equation}
    \hat{H}^{[N]}_\mathrm{LM}=\hbar g_\mathrm{c} \sqrt{\omega_{\bf k}} \cdot\sum_{j}\sum_{\bf k} (\hat{\sigma}_j + \hat{\sigma}^{\dagger}_j) (\hat{a}_{\bf k} + \hat{a}^{\dagger}_{\bf k}),
\end{equation}
where $g_\mathrm{c} = \mu_\mathrm{LL'} \lambda_\mathrm{c} / \sqrt{2} \hbar$. When assuming $\cos\varphi_{j}=1$, Eq.~\ref{GenPF_LWA} will have permutation symmetry, and one can introduce the collective operators 
\begin{equation}
    \hat{\sigma}_{N}^\dagger=\frac{1}{\sqrt{N}}\sum_{j}|\nu_{j}\rangle\langle G|,~~\hat{\sigma}_{N}=\frac{1}{\sqrt{N}}\sum_{j}|G\rangle\langle \nu_{j}|.
\end{equation}
Using the collective operators, and ignoring the counter-rotating wave term ($\propto\sum_{j,{\bf k}}(\hat{\sigma}^{\dagger}_{j}\hat{a}^{\dagger}_{\bf k}+\hat{\sigma}_{j}\hat{a}_{\bf k})$), which is less important for the resonant condition of $\omega_{\bf k}=\omega_{0}$ when $g_\mathrm{c}$ is small, we have
\begin{equation}
    \hat{H}^{[N]}_\mathrm{LM}=\sqrt{N} \hbar g_\mathrm{c}\sqrt{\omega_{\bf k}} \cdot\sum_{\mathrm k}(\sigma^{\dagger}_{N}\hat{a}_{\bf k}+\sigma_{N}\hat{a}^{\dagger}_{\bf k}).
\end{equation}
This level of approximation is commonly referred to as the Holstein-Tavis-Cummings model. Note that the effective coupling between light and matter is now $\sqrt{N}g_\mathrm{c}$. This light-matter coupling term will hybridize the 1 photon-dressed ground state $|G\rangle\otimes|1_{\bf 
 k}\rangle$ with the 0-photon dressed bright state $|B\rangle|\otimes|0_{\bf k}\rangle$, generating the following polariton states
\begin{subequations}
\begin{align}\label{eq:TC-polariton}
    |+\rangle &= \cos\phi_{N}\cdot|\mathrm{B}\rangle\otimes |0_{\bf k}\rangle  + \sin\phi_{N} \cdot |G\rangle\otimes|1_{\bf k}\rangle \\
    |-\rangle &= -\sin\phi_{N}\cdot|\mathrm{B}\rangle\otimes|0_{\bf k}\rangle + \cos\phi_{N}\cdot |G\rangle \otimes|1_{\bf k}\rangle,
\end{align}
\end{subequations}
where the mixing angle is 
\begin{equation}
    \phi_{N} =\frac{1}{2}\tan^{-1}[ (2\sqrt{N \omega_{\bf k}}g_\mathrm{c}) / (\omega_{\bf k} (k_{\parallel}) - \omega_{0})],
\end{equation}
with the maximum mixing between light and matter being reached when $\omega_\mathrm{c} ({\bf k}) = \omega_{0}$. The dark states, on the other hand, do not mix with the photonic DOF under this simplified approximation and remain to be $|\mathrm{D}_{\alpha}\rangle\otimes|0_{\bf k}\rangle$ in the singly excited subspace. It is a well-known fact  that $\langle G|\otimes\langle 1_{\bf k}|\hat{H}_\mathrm{LM}|\mathrm{D}_{\alpha}\rangle\otimes|0_{\bf k}\rangle=0$, because $\langle G|\sum_{j}(\sigma_{j}+\sigma^{\dagger}_{j})|\sum_{k}\mathcal{C}^{\alpha}_{k}|\nu_{k}\rangle=\sum_{j}\langle \nu_{j}|\sum_{k}\mathcal{C}^{\alpha}_{k}|\nu_{k}\rangle=\sum_{k}\mathcal{C}^{\alpha}_{k}=0$. The energy gap between the upper polariton state and the lower polariton state is referred to as the Rabi splitting and is expressed as follows
\begin{equation}\label{wR}
    \Omega_\mathrm{R}\equiv E_{+}-E_{-} = \sqrt{(\omega_{\bf k}-\omega_{0})^2 + 4 N \omega_\mathrm{c} g^2_\mathrm{c}},
\end{equation}
and under the resonant condition $\omega_{\bf k}=\omega_{0}$, the Rabi splitting is $\Omega_{R}=2\sqrt{N \omega_\mathrm{c}}g_\mathrm{c}=\sqrt{\frac{2\omega_{\bf k}}{\epsilon_{0}\hbar}}\sqrt{\frac{N}{\mathcal V}}\cdot {\mu_\mathrm{LL'}}$. As one can clearly see, forming the Rabi splitting is originated from a collective phenomenon, resulting in the well know $\sqrt{N}$ dependence or $\sqrt{N/\mathcal{V}}$ dependence (square root of concentration), confirmed by experiments~\cite{Ebbesen_nanophotonics_2020}.

The dark states, on the other hand, do not mix with the photonic DOF under this simplified approximation and remain to be $|\mathrm{D}_{\alpha}\rangle\otimes|0_{\bf k}\rangle$ in the single excited subspace. This is because that $\langle G|\otimes\langle 1_{\bf k}|\hat{H}_\mathrm{LM}|\mathrm{D}_{\alpha}\rangle\otimes|0_{\bf k}\rangle=0$, because $\langle G|\sum_{j}(\sigma_{j}+\sigma^{\dagger}_{j})|\sum_{k}\mathcal{C}^{\alpha}_{k}|\nu_{k}\rangle=\sum_{j}\langle \nu_{j}|\sum_{k}\mathcal{C}^{\alpha}_{k}|\nu_{k}\rangle=\sum_{k}\mathcal{C}^{\alpha}_{k}=0$. As such, these states are \textit{dark} because they do not contain photonic components, but also there is no optical transition to them from $|G\rangle$.

\section{IV. The effective Hamiltonian and spectral density}
Recall the total Hamiltonian reads as (c.f. Eq.~\ref{GenPF_LWA}) 
\begin{align} \label{H_tot}
    \hat{H} &= \sum_{j = 1}^N \Big[\frac{\hat{P}^2_j}{2M} + V(\hat{R_j})\Big] + \hat{H}_{\nu} + \frac{1}{2} \sum_{\mathbf{k}} \Big[ \hat{p}^2_\mathrm{\mathbf{k}} + \omega^2_\mathbf{k} \Big( \hat{q}_\mathrm{\mathbf{k}} + {\frac{\lambda_\mathrm{c}}{\omega_{\bf k}}}  \sum_{j=1}^{N} \mu(\hat{R}_{j})\cdot\cos\varphi_{j} \Big)^2 \Big] + \hat{H}_\mathrm{loss}.
\end{align}
It is shown that the the model Hamiltonian has a one-to-one map (through normal mode transformation) to the effective Hamiltonian as below,~\cite{Ambegaokar_1985, Miller_2001}
\begin{align} \label{H_mapping}
    \hat{H} = \sum_{j = 1}^N \Big[\frac{\hat{P}^2_j}{2M} + V(\hat{R_j})\Big] + \hat{H}_{\nu} + \frac{1}{2} \sum_{\mathbf{k}, \zeta} \Big[ \hat{\tilde{p}}^2_j + \tilde{\omega}^2_{\mathbf{k}, \zeta} \Big(\hat{\tilde{x}}_{\mathbf{k}, \zeta} - \frac{\tilde{c}_{\mathbf{k}, \zeta}}{\tilde{\omega}^2_{\mathbf{k}, \zeta}} \sum_{j=1}^{N} \mu(\hat{R}_{j})\cdot\cos\varphi_{j} \Big)^2 \Big],
\end{align}
where the effective bath and its interaction with the system dissipation modes are described by an effective spectral density function,
\begin{align} \label{J_eff_prelude}
    J_{\mathrm{eff}}(\omega) \equiv \frac{\pi}{2} \sum_{\mathbf{k}, \zeta} \frac{\tilde{c}^2_{\mathbf{k}, \zeta}}{\tilde{\omega}_{\mathbf{k}, \zeta}} \delta(\omega - \tilde{\omega}_{\mathbf{k}, \zeta}).
\end{align}
Note that the effective bath serves as the common bath for all the system DOFs. Now we provide a concise justification for this conclusion, and derive the explicit expression of the effective spectral density function. For convenience, we denote $\mathcal{\hat{S}} = \sum_{j=1}^{N} \mu(\hat{R}_{j})\cdot\cos\varphi_{j}$ as the collective system dipole operator (This is equivalent to the $\mathcal{\hat{S}}_{\bf k}$ from Eq.~\ref{eq:SK} when under the long-wavelength approximation).
The terms in the total Hamiltonian (Eq.~\ref{H_tot}) that of interest for this normal mode transformation read as 
\begin{align} \label{Ham1}
    \hat{H} - \hat{H}_\nu = \hat{H}'=&\sum_{j = 1}^N \left[\frac{\hat{P}^2_j}{2M} + V(\hat{R_j})\right] + \frac{1}{2} \sum_{\mathbf{k}} \left[ \hat{p}^2_\mathrm{\mathbf{k}} + \omega^2_\mathbf{k} \left( \hat{q}_\mathrm{\mathbf{k}} + \frac{\lambda_\mathrm{c}}{\omega_\mathbf{k}} \hat{\mathcal{S}} \right)^2 \right] + \frac{1}{2} \sum_{\mathbf{k}, \zeta} \left[ \hat{\tilde{p}}^2_{\mathbf{k}, \zeta} + \tilde{\omega}^2_{\mathbf{k}, \zeta} \left(\hat{\tilde{x}}_{\mathbf{k}, \zeta} - \frac{\tilde{c}_{\mathbf{k}, \zeta}}{\tilde{\omega}^2_{\mathbf{k}, \zeta}} \hat{q}_\mathrm{\mathbf{k}} \right)^2 \right],
\end{align}
where $\hat{H}_{\nu}$ is omitted for not being directly involved in light-matter interactions. Next, by applying harmonic analysis to the equations of motion, we derive the effective spectral density function which describes the cavity modes as well as their associated loss. We will follow and generalize the approach proposed by Leggett~\cite{Leggett_1984} and Garg, et al.~\cite{Ambegaokar_1985}. The classical equations of motion with respect to the Hamiltonian in Eq.~\ref{Ham1} can be formally written down as
\begin{subequations} \label{classical_EOMs}
\begin{align}
    & M_\mathrm{eff} \ddot{\mathcal{S}} = - \frac{\partial V(\{R_j\})}{\partial \mathcal{S}} - \sum_{\mathbf{k}} \omega^2_\mathbf{k} \cdot \frac{\lambda_\mathrm{c}}{\omega_\mathbf{k}} \left(q_\mathrm{\mathbf{k}} + \frac{\lambda_\mathrm{c}}{\omega_\mathbf{k}} \mathcal{S} \right),\\
    & \ddot{q}_\mathrm{\mathbf{k}} = - \omega^2_\mathbf{k} \left(q_\mathrm{\mathbf{k}} + \frac{\lambda_\mathrm{c}}{\omega_\mathbf{k}} \mathcal{S} \right) + \sum_{\zeta} \left( \tilde{c}_{\mathbf{k}, \zeta} \tilde{x}_{\mathbf{k}, \zeta} - \frac{ \tilde{c}^2_{\mathbf{k}, \zeta}}{\tilde{\omega}^2_{\mathbf{k}, \zeta}} q_\mathrm{\mathbf{k}} \right), \\
    & \ddot{\tilde{x}}_{\mathbf{k}, \zeta} = - \tilde{\omega}^2_{\mathbf{k}, \zeta} \tilde{x}_{\mathbf{k}, \zeta} + \tilde{c}_{\mathbf{k}, \zeta} q_\mathrm{\mathbf{k}} .
\end{align}
\end{subequations}
where $M_\mathrm{eff}$ is the effective mass. Applying Fourier transform to Eq.~\ref{classical_EOMs} leads to
\begin{subequations}
    \begin{align}
    & \left(- M_\mathrm{eff} \omega^2 + \sum_{\mathbf{k}} \lambda^2_{\mathrm{c}} \right) \mathcal{S}(\omega) + \sum_{\mathbf{k}} \lambda_{\mathrm{c}} \omega_{\mathbf{k}} q_\mathrm{\mathbf{k}}(\omega) = - V'_{\omega}, \label{FT_1} \\
    & \left[( \omega_\mathbf{k}^2 - \omega^2 ) + \sum_j \frac{\tilde{c}^2_{\mathbf{k}, \zeta}}{\tilde{\omega}^2_{\mathbf{k}, \zeta}} \right] q_\mathrm{\mathbf{k}}(\omega) - \sum_\zeta \tilde{c}_{\mathbf{k}, \zeta} \tilde{x}_{\mathbf{k}, \zeta}(\omega) + \lambda_{\mathrm{c}} \omega_{\mathbf{k}} \mathcal{S}(\omega) = 0, \label{FT_3} \\
    & (- \omega^2 + \tilde{\omega}^2_{\mathbf{k}, \zeta}) \tilde{x}_{\mathbf{k}, \zeta}(\omega) - \tilde{c}_{\mathbf{k}, \zeta} q_\mathrm{\mathbf{k}}(\omega) = 0, \label{FT_4}
\end{align}
\end{subequations}
where $V'_{\omega}$ is the Fourier transform of $\partial V(\{R_j\}) / \partial \mathcal{S}$. Plugging Eq.~\ref{FT_4} into \ref{FT_3} to cancel the $\tilde{x}_{\mathbf{k}, \zeta}(\omega)$ terms, one obtains
\begin{align}\label{qc_w_1}
    \left[ \omega_\mathbf{k}^2 - \omega^2 \left( 1 + \sum_\zeta \frac{\tilde{c}^2_{\mathbf{k}, \zeta}}{\tilde{\omega}^2_{\mathbf{k}, \zeta} (- \omega^2 + \tilde{\omega}^2_{\mathbf{k}, \zeta})} \right) \right] q_\mathrm{\mathbf{k}}(\omega) + \lambda_{\mathrm{c}} \omega_{\mathbf{k}} \mathcal{S}(\omega) = 0.
\end{align}
We further define 
\begin{align}\label{Lw_def}
    L_{\mathbf{k}}(\omega) = - \omega^2 \left[ 1 + \sum_\zeta \frac{\tilde{c}^2_{\mathbf{k}, \zeta}}{\tilde{\omega}^2_{\mathbf{k}, \zeta} (- \omega^2 + \tilde{\omega}^2_{\mathbf{k}, \zeta})} \right],
\end{align}
Eq.~\ref{qc_w_1} becomes
\begin{align}\label{qc_FT}
    q_\mathrm{\mathbf{k}}(\omega) = - \frac{\lambda_{\mathrm{c}} \omega_{\mathbf{k}} \mathcal{S}(\omega)}{\omega_\mathbf{k}^2 + L_{\mathbf{k}}(\omega)}.
\end{align}
Notice that $L_{\mathbf{k}}(\omega)$ can be alternatively expressed as
\begin{align}\label{Lw_rewrite}
    L_{\mathbf{k}}(\omega) =  - \omega^2 \left[ 1 + \int_{0}^{+ \infty} ds \frac{\sum_\zeta \frac{\tilde{c}^2_{\mathbf{k}, \zeta}}{\tilde{\omega}_{\mathbf{k}, \zeta}} \delta(s - \tilde{\omega}_{\mathbf{k}, \zeta})}{s (s^2 - \omega^2)} \right] = - \omega^2 \left[ 1 + \frac{2}{\pi} \int_{0}^{+ \infty} ds \frac{J_{\mathrm{loss}}(s, \mathbf{k})}{s (s^2 - \omega^2)} \right],
\end{align}
where we used Eq.~\ref{spectral_density_loss} to give rise to the loss spectral density function. Plugging Eq.~\ref{qc_FT} into \ref{FT_1}, one obtains
\begin{align}\label{Kw_def}
    K(\omega) \mathcal{S}(\omega) \equiv \left(- M_\mathrm{eff} \omega^2 + \sum_{\mathbf{k}} \frac{\lambda^2_{\mathrm{c}} L_{\mathbf{k}}(\omega)}{\omega_\mathbf{k}^2 + L_{\mathbf{k}}(\omega)} \right) \mathcal{S}(\omega) = - V'_{\omega},
\end{align}
And the total spectral density function felt by the reaction coordinate $\mathcal{S}$ is given by the branch cut of $K(z)$ on the complex plane, $J(\omega) = \lim_{\epsilon \rightarrow 0^+} \mathrm{Im}[K(\omega - i\epsilon)]$. For our Eq.~\ref{Kw_def}, it reads as~\cite{Miller_2001, Manolopoulos_2019}
\begin{align}\label{Jeff}
    J_{\mathrm{eff}}(\omega) \equiv \frac{\pi}{2} \sum_{\mathbf{k}, \zeta} \frac{\tilde{c}^2_{\mathbf{k}, \zeta}}{\tilde{\omega}_{\mathbf{k}, \zeta}} \delta(\omega - \tilde{\omega}_{\mathbf{k}, \zeta}) = \sum_{\mathbf{k}} \frac{ \lambda^2_{\mathrm{c}} \omega^2_\mathbf{k} J_{\mathrm{loss}}(\omega, \mathbf{k})}{\left[ \omega^2_\mathbf{k} - \omega^2 + \xi_{\mathbf{k}}(\omega) \right]^2 + [J_{\mathrm{loss}}(\omega, \mathbf{k})]^2},
\end{align}
where $\xi_{\mathbf{k}}(\omega)$ is expressed as
\begin{align}\label{xi_w}
    \xi_{\mathbf{k}}(\omega) = \frac{2\omega^2}{\pi} \mathcal{P} \int_0^{\infty} ds\ \frac{J_{\mathrm{loss}}(s, \mathbf{k})}{s(\omega^2 - s^2)}.
\end{align}
And $\mathcal{P}$ in the above expression denotes to principal value integral. Notice that Eq.~\ref{Jeff} is additive with respect to $\mathbf{k}$, meaning the total spectral density function is contributed by all of the cavity modes as well as their associated loss. Further, the normal mode transformations are restricted by the following identities,~\cite{Miller_2001}
\begin{align}
    \lambda_\mathrm{c} \omega_\mathbf{k} \hat{q}_\mathbf{k} = \sum_{\zeta} \tilde{c}_{{\bf k}, \zeta} \hat{\tilde{x}}_{{\bf k}, \zeta},~~~~~ \sum_\zeta \tilde{c}^2_{{\bf k}, \zeta} = \lambda_\mathrm{c}^2 \omega_\mathbf{k}^2,~~~~~ \omega_\mathbf{k} = \lambda_\mathrm{c}^2 \omega_\mathbf{k}^2 \cdot \left(\sum_\zeta \tilde{c}^2_{{\bf k}, \zeta} / \tilde{\omega}^2_{{\bf k}, \zeta} \right)^{-1}.
\end{align}
To simplify our argument, we assume that the cavity loss is homogeneous (\textit{i.e.}, does not depend on $\mathbf{k}$) and strictly Ohmic (\textit{i.e.}, Markovian), which means 
\begin{align}
    J_{\mathrm{loss}}(\omega, \mathbf{k}) = \alpha \omega \exp(- \omega/\omega_\mathrm{m}),~~~ \omega_\mathrm{m} \rightarrow + \infty,
\end{align}
where $\alpha \equiv \tau^{-1}_\mathrm{c}$ is the inverse of the cavity lifetime $\tau_\mathrm{c}$. Under the Markovian limit, $\xi_{\mathbf{k}}(\omega) \rightarrow 0$, Eq.~\ref{Jeff} is simplified as
\begin{align} \label{Jeff_Markovian_k}
    J_{\mathrm{eff}}(\omega) = \sum_{\mathbf{k}} \frac{ \lambda^2_{\mathrm{c}} \omega^2_\mathbf{k}  \tau^{-1}_\mathrm{c} \omega}{\left( \omega^2_\mathbf{k} - \omega^2 \right)^2 + \tau^{-2}_\mathrm{c} \omega^2},
\end{align}
which is Eq.~11 of the main text.

\section{V. Density of States, and the effective spectral density}
The cavity dispersion relation is written as (c.f. Eq.~1 of the main text)
\begin{align} \label{dispersion}
    \omega_{\mathbf{k}} = \frac{c}{n_\mathrm{c}}\sqrt{k^2_\perp + k^2_{\parallel}} = \frac{ck_\perp}{n_\mathrm{c}}\sqrt{1 + \tan^2 \theta}.
\end{align}
where $\tan \theta= k_{\parallel}/k_{\perp}$. In turn, one can calculate $k_{\parallel}$ in terms of the cavity frequency $\omega$ as 
\begin{align}
    k_{\parallel} = \pm \frac{n_\mathrm{c} \sqrt{\omega_{\bf k}^2 - \omega^2_\mathrm{c}}}{c},
\end{align}
where for $k_{\parallel}=0$ we introduce 
\begin{equation}
    \omega_{\mathrm{c}} = ck_\perp / n_\mathrm{c}
\end{equation}
which is the photon frequency associated with the quantized direction (normal incidence). Evaluating the derivative of $k_{\parallel}$ with respect to $\omega$, one obtains
\begin{align}
    \frac{dk_{\parallel}}{d\omega_{\bf k}} = \pm \frac{n_\mathrm{c} \omega_{\bf k}}{c \sqrt{\omega_{\bf k}^2 - \omega^2_\mathrm{c}}}.
\end{align}
The density of states (DOS) is defined as
\begin{align}\label{DOS_w}
    g(\omega) = \frac{c}{2 n_\mathrm{c} k^\mathrm{m}_{\parallel}} \int_{- k^\mathrm{m}_{\parallel}}^{+ k^\mathrm{m}_{\parallel}} \delta(\omega_{\mathbf{k}}-\omega)\ dk_{\parallel} &= \frac{c}{2 n_\mathrm{c} k^\mathrm{m}_{\parallel}} \int_0^{\omega_\mathrm{m} } \delta(\omega_{\mathbf{k}}-\omega) \frac{dk_{\parallel}}{d\omega_{\mathbf{k}}} d\omega_{\mathbf{k}} = \frac{\omega}{ k^\mathrm{m}_{\parallel} \sqrt{\omega^2 - \omega^2_c}} \cdot \Theta(\omega-\omega_{c}),
\end{align}
where $k^\mathrm{m}_{\parallel}$ is the cutoff value of the in-plane wavevector, $\omega_\mathrm{m} = \frac{c}{n_\mathrm{c}}\sqrt{k^2_\perp + (k^\mathrm{m}_{\parallel})^2} \to \infty$ is the cutoff frequency, and $\Theta(\omega-\omega_\mathrm{c})$ is the Heaviside function, such that $\Theta(\omega-\omega_\mathrm{c})=1$ when $\omega\ge\omega_\mathrm{c}$ and $\Theta(\omega-\omega_\mathrm{c})=0$ when $\omega<\omega_\mathrm{c}$, such that there is no mode density below $\omega_{c}$. Note that there is a singularity in the DOS at $\omega = \omega_{\mathrm{c}}$, which could play a crucial role in explaining the $k_{\parallel}$ dependence of VSC. Note that Eq.~\ref{DOS_w} can be equivalently expressed as
\begin{align}\label{DOS_k}
g(k_{\parallel}) =\frac{1}{k^\mathrm{m}_{\parallel}} \frac{\sqrt{k^2_{\parallel}+k^2_{\perp}}}{k_{\parallel}}\cdot \Theta(\omega-\omega_{c})= \frac{1}{2 k^\mathrm{m}_{\parallel}} \sqrt{1 + \left(\frac{n_\mathrm{c} \omega_{\mathrm{c}}}{c k_{\parallel}} \right)^2} ,~~~ - k^\mathrm{m}_{\parallel} \leq k_{\parallel} \leq k^\mathrm{m}_{\parallel},
\end{align}
where under the limit of the $k^\mathrm{m}_{\parallel} \to \infty$, the above density of state reduces to
\begin{equation} \label{DOS_delta}
    \lim_{k^\mathrm{m}_{\parallel} \to \infty} g(k_{\parallel})\approx \delta(k_{\parallel}).
\end{equation}
One can easily check the normalization of $g(k_{\parallel})$, which is $\int_{-k^\mathrm{m}_{\parallel}}^{k^\mathrm{m}_{\parallel}} dk_{\parallel}~ g(k_{\parallel}) = 1$. Equivalently, by defining the maximal incident angle $\theta_{\mathrm{m}}$ via $k^\mathrm{m}_{\parallel} = k_{\perp} \tan \theta_{\mathrm{m}}$, and with ${n_\mathrm{c} \omega_{\mathrm{c}}}/{c k_{\parallel}}=k_{\perp}/k_{\parallel}=\cot\theta$, the density of state $g$ can be expressed as a function of the incident angle $\theta$ as follows
\begin{align}\label{DOS_theta}
    g(\theta) = \frac{1}{2 k_\perp \tan \theta_{\mathrm{m}}} \sqrt{1 + \cot^2 \theta},~~~  (\mathrm{for}~-\theta_{\mathrm{m}} \leq \theta \leq \theta_{\mathrm{m}}).
\end{align}
It is easy to check that the normalization conditions 
\begin{align*}
    \int_{0}^{\omega_\mathrm{m}} d\omega~ (dk_{\parallel} / d \omega) g(\omega) = 1,~~~~~ \int_{-\theta_{\mathrm{m}}}^{\theta_{\mathrm{m}}} d\theta~ (dk_{\parallel} / d\theta) g(\theta) = 1
\end{align*}
are both satisfied. The singularity of $g$ appears at $k_{\parallel} = 0$ or $\theta = 0$, but the integral of it is finite and well-defined.

For a quasi-continuous wave vector $\mathbf{k}$, we can rewrite the summation in Eq.~\ref{Jeff} into integration. The integration measure should be chosen as $d k_{\parallel}$ as it is physically expected, which means varying $k_{\parallel}$ evenly. Here we will convert it to $d\theta$ via chain rule, and every $\mathbf{k}$-dependent term is expressed in terms of $\theta$. Recall that we derived the DOS of the Fabry-P\'erot cavity as (c.f. Eq.~\ref{DOS_theta})
\begin{align*}
    g(\theta) = \frac{1}{2 k_\perp \tan \theta_{\mathrm{m}}} \sqrt{1 + \cot^2 \theta},~~~ - \theta_{\mathrm{m}} \leq \theta \leq \theta_{\mathrm{m}},~~~ \theta_{\mathrm{m}} \to \frac{\pi}{2},
\end{align*}
which will reduce to a $\delta$-function $\delta(\theta) / k_\perp $ when $\theta_\mathrm{m} \to \pi / 2$. And the dispersion relation is (c.f. Eq.~\ref{dispersion})
\begin{align*}
    k_{\parallel} = k_\perp \tan \theta,~~~~~~ \omega_{\mathbf{k}} = \omega_\mathrm{c} \sqrt{1 + \tan^2 \theta}.
\end{align*}
As a result, Eq.~\ref{Jeff} becomes
\begin{align} \label{Jeff_nonMarkov}
J_{\mathrm{eff}}(\omega) &= \int_{- k^\mathrm{m}_{\parallel}}^{k^\mathrm{m}_{\parallel}} d k_{\parallel}~ g(\theta) \frac{ \lambda^2_{\mathrm{c}} \omega^2_\mathbf{k}  \tau^{-1}_\mathrm{c} \omega}{\left( \omega^2_\mathbf{k} - \omega^2 \right)^2 + \tau^{-2}_\mathrm{c} \omega^2} = \frac{\omega_{\mathrm{c}}^2 \lambda^2_{\mathrm{c}}}{2 k_\perp \tan \theta_{\mathrm{m}}} \int_{- \theta_{\mathrm{m}}}^{\theta_{\mathrm{m}}} d\theta~ \frac{dk_{\parallel}}{d \theta} \times \sqrt{1 + \cot^2 \theta} \times \frac{ (1 + \tan^2 \theta) \tau^{-1}_\mathrm{c} \omega}{\left( \omega^2_\mathbf{k} - \omega^2 \right)^2 + \tau^{-2}_\mathrm{c} \omega^2} \notag\\
    &= \frac{ \omega^2_{\mathrm{c}} \lambda^2_{\mathrm{c}} }{\tan \theta_{\mathrm{m}}} \int_{0}^{\theta_{\mathrm{m}}} d\theta\ \frac{\csc \theta}{\cos^4 \theta} \times \frac{\tau^{-1}_\mathrm{c} \omega}{\left( \omega^2_\mathbf{k} - \omega^2 \right)^2 + \tau^{-2}_\mathrm{c} \omega^2},
\end{align}
which is Eq.~12 of the main text. Here, we take the limit of $\theta_\mathrm{m} \to \pi / 2$, so that the DOS will reduce to $\delta(\theta) / k_\perp$ (see Eq.~\ref{DOS_k}). As a result, one can obtain from Eq.~\ref{Jeff_nonMarkov} that
\begin{align} \label{Jeff_nmk_single}
    J_{\mathrm{eff}}(\omega) &= {\omega^2_{\mathrm{c}} \lambda^2_{\mathrm{c}} } \cdot \frac{\tau^{-1}_\mathrm{c} \omega}{\left( \omega^2_\mathrm{c} - \omega^2 \right)^2 + \tau^{-2}_\mathrm{c} \omega^2},
\end{align}
which is Eq.~13 of the main text. Note that Eq.~\ref{Jeff_nmk_single} reduces to the single mode case. The validation of single-mode approximation in the VSC problem will greatly benefit us to simplify the analysis. Physically, it means that the DOS at $\theta \neq 0$ is infinitesimal comparing to the DOS at $\theta = 0$. But this is not in contrary to the fact that there is still Rabi-splitting when $\omega_0 > \omega_\mathrm{c}$, because there are indeed modes (and associated photonic states) at that higher frequency. 

An alternative way to understand the result in Eq.~\ref{Jeff_nmk_single} is that we can separate the singular point at $\theta = 0$,
\begin{align}
    \lim_{\theta_\mathrm{m} \rightarrow \frac{\pi}{2}} \int_{0}^{\theta_{\mathrm{m}}} f(\theta) d\theta = \lim_{\delta \rightarrow 0,~ \theta_\mathrm{m} \rightarrow \frac{\pi}{2}} \left(\int_{0}^{\delta}f(\theta) d\theta + \int_{\delta}^{\theta_{\mathrm{m}}} f(\theta) d\theta \right) = \lim_{\delta\to0}f(\theta=\delta) + \lim_{\delta \rightarrow 0,~ \theta_\mathrm{m} \rightarrow \frac{\pi}{2}} \int_{\delta}^{\theta_{\mathrm{m}}}f(\theta) d\theta,
\end{align}
such that 
\begin{align}
    \lim_{\theta_\mathrm{m} \rightarrow \frac{\pi}{2}} \frac{1}{\tan \theta_{\mathrm{m}}} \int_{0}^{\theta_{\mathrm{m}}} d\theta\ \frac{\csc \theta}{\cos^4 \theta} = \lim_{\delta \rightarrow 0,~ \theta_\mathrm{m} \rightarrow \frac{\pi}{2}} \frac{\csc \delta}{\tan \theta_{\mathrm{m}}} + \mathcal{O}(\delta) = 1,
\end{align}
where we take $\delta$ and $\frac{\pi}{2} 
       - \theta_\mathrm{m}$ to have the same order of magnitude, and $\lim_{\delta \rightarrow 0,~ \theta_\mathrm{m} \rightarrow \frac{\pi}{2}} \frac{\csc \delta}{\tan \theta_{\mathrm{m}}} = 1$ as is required by the normalization condition of the DOS. As such, Eq.~\ref{Jeff_nmk_single} is recovered. 

\begin{figure}[htbp]
    \centering
    \includegraphics[width=0.8\linewidth]{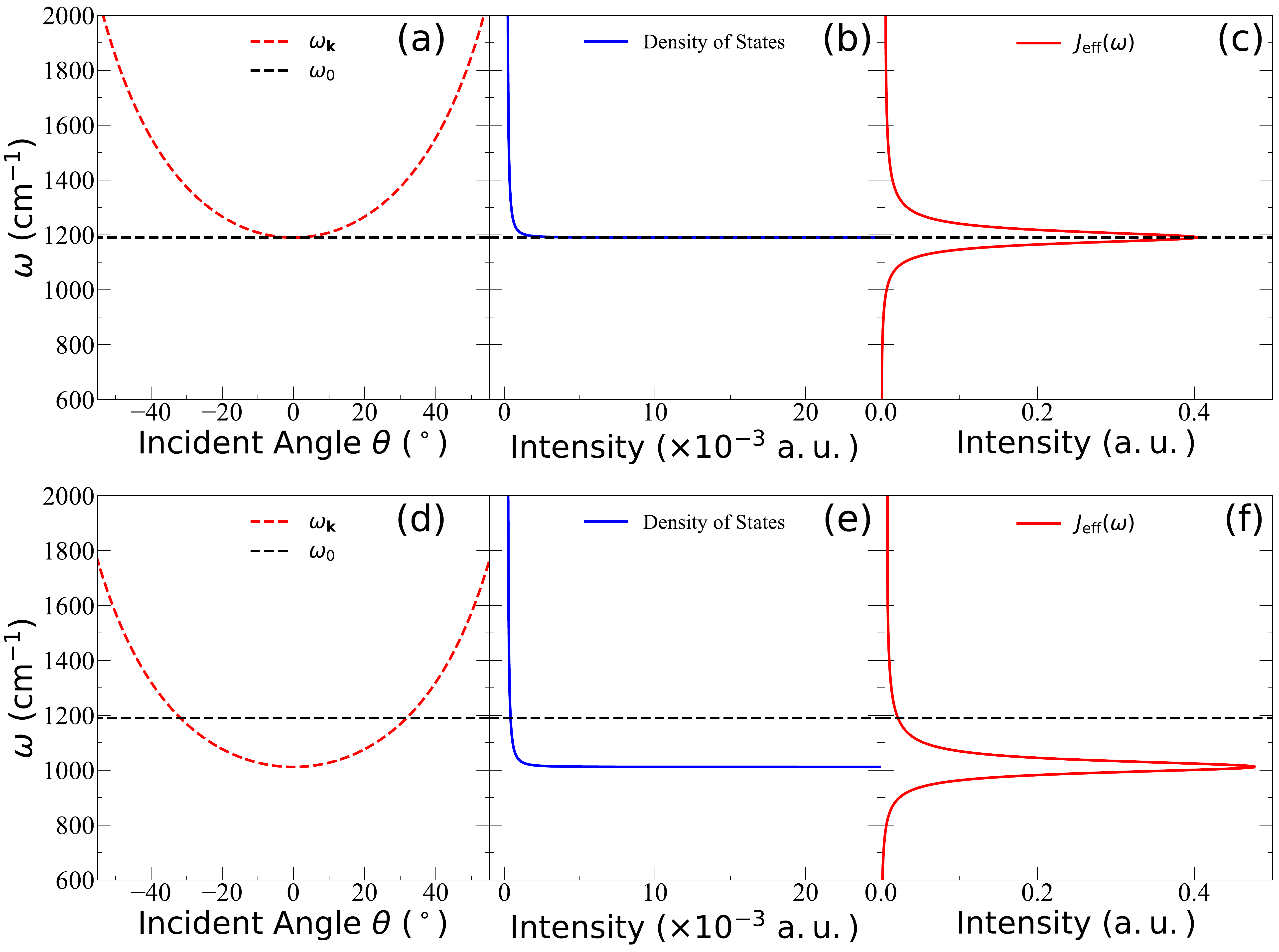}
    \caption{(a) The cavity dispersion relation, (b) density of states, and (c) effective spectral density function for the normal incidence case (with $\omega_\mathrm{c} = \omega_0$), where the resonance condition is reached at $\theta = 0$. (d)-(f) correspond to the red-detuned case (oblique incidence), with $\omega_\mathrm{c} = 0.85~ \omega_0$, whose resonance condition is reached at around $\theta = 32^\circ$. The broadening parameter is taken as $\alpha = 50~ \mathrm{cm}^{-1}$, which corresponds to a cavity lifetime of $\tau_\mathrm{c} \approx 106~ \mathrm{fs}$.}
    \label{fig_DOS}
\end{figure}

Fig.~\ref{fig_DOS} presents the cavity dispersion relation, density of states, and the effective spectral density function. One can clearly see that under the normal incident condition  in panels (a)-(c) when $\theta = 0$, $J_\mathrm{eff}(\omega_0)$ is maximized at $\omega_\mathrm{c} = \omega_0$, as will the rate constant based on the FGR expression (Eq.~15 of the main text). While at the detuned cases ($\omega_\mathrm{c} \neq \omega_0$ or $|\theta| > 0$), the intensity of $J_\mathrm{eff}(\omega_0)$ is diminished. This explains the experimentally observed resonance phenomena which only occur at $\omega_\mathrm{c}=\omega_{0}$ when $k_{\parallel}=0$.

\section{VI. Derivation of Eq.~20 of the main text}
Here, we theoretically justify Eq.~20 of the main text, for the general case of $\tau_\mathrm{c}$ dependence. In the case of a very large $\tau_\mathrm{c}$, the main term that causes the broadening of the $|\nu^{j}_\mathrm{L}\rangle\to|\nu'^{j}_\mathrm{L}\rangle$ transition comes from the $\hat{H}_{\nu}$ term in Eq.~\ref{Hnu}, which must then be considered in the rate expression. This is a Holstein-type system-bath coupling (diagonal coupling) and is identical for all molecules $j\in[1,N]$, which is expressed as
\begin{equation}\label{system-bath}
    \Big[R_{\mathrm{LL}}|\nu^{j}_\mathrm{L}\rangle \langle \nu^{j}_\mathrm{L}| + R_{\mathrm{L}'\mathrm{L}'} |\nu'^{j}_\mathrm{L}\rangle \langle \nu'^{j}_\mathrm{L}|\Big] \otimes \hat{F}^{j}_\nu\equiv \epsilon_{z}\cdot\frac{1}{2}\sigma^j_{z}\otimes \hat{F}^{j}_\nu,
\end{equation}
where $R_{\mathrm{LL}}=\langle \nu^{j}_\mathrm{L}|\hat{R}_{j}|\nu^{j}_\mathrm{L}\rangle$, similar for $R_{\mathrm{L'L}'}$, and the stochastic force operator of the phonon is
\begin{align}
    \hat{F}^{j}_\nu = \sum_\zeta c_\zeta \hat{x}_{j,\zeta}.
\end{align}
Further, $\hat{\sigma}^j_z = |\nu'^j_\mathrm{L}\rangle \langle \nu'^j_\mathrm{L}| - |\nu^j_\mathrm{L}\rangle \langle \nu^j_\mathrm{L}|$, and $\epsilon_{z}=R_{\mathrm{LL}}-R_{\mathrm{L'L'}}=0.232$ a.u. for the model system considered in Sec.~II. Eq.~\ref{system-bath} is mainly responsible for the inhomogeneous broadening effect in the spectra, and it should also broaden the transition frequency $\omega_{0}$. The variance of this fluctuation term is~\cite{Ratner_2003, Huo_2017, Sutirtha_2021}
\begin{equation}\label{sigma}
    \sigma^2 = \epsilon^2_{z}\cdot \frac{1}{\pi} \int_{0}^\infty d\omega~ J_\nu (\omega) \coth(\beta \omega / 2),
\end{equation}
where $J_\nu (\omega)$ is identical for all molecules. With the above analysis, the rate constant in a general way can be written down as the following convolution expression~\cite{Mukamel}
\begin{equation}\label{kvsc-int}
    \tilde{k}_\mathrm{VSC}= \int_{0}^{\infty} d\omega~ k_\mathrm{VSC}(\omega) G(\omega-\omega_{0}),
\end{equation}
where the broadening function $G(\omega-\omega_{0})$ is a Gaussian distribution centered around $\omega_0$, defined as
\begin{align}\label{gaussian}
    G(\omega-\omega_{0}) &= \frac{1}{\sqrt{2\pi \sigma^2}} \exp\left[ - \frac{(\omega - \omega_0)^2}{2 \sigma^2} \right],
\end{align}
with the width expressed in Eq.~\ref{sigma}. This is the result of Eq.~20 of the main text.

There are several interesting limits of the expression in Eq.~\ref{kvsc-int}. First, in the lossless limit ($\tau_\mathrm{c} \rightarrow \infty$) and the effective spectral density function in Eq.~\ref{Jeff_nmk_single} (Eq. 13 of the main text) will reduce to a single $\delta$-function, $J_\mathrm{eff}(\omega) \approx \frac{\pi}{2} \lambda^2_\mathrm{c} \omega_\mathrm{c} \delta(\omega - \omega_\mathrm{c})$. As a result, the broadening is fully dictated by the variance of the Gaussian. If we focus on the case of fully isotropic dipoles, then 
\begin{equation}\label{kvsc_lossless}
    \tilde{k}_\mathrm{VSC} \approx \frac{2\pi}{3}\cdot N g^2_\mathrm{c} \omega_\mathrm{c} n(\omega_0) \cdot\int_{0}^{\infty} d\omega~ \delta(\omega - \omega_\mathrm{c}) G(\omega-\omega_{0}) = \frac{2\pi}{3}\cdot N g^2_\mathrm{c} \omega_\mathrm{c}\cdot G(\omega_\mathrm{c} - \omega_0)\cdot e^{-\beta\hbar\omega_{0}}
\end{equation}

The rate profile described in Eq.~\ref{kvsc_lossless} is a Gaussian function centered at $\omega_0$ with respect to cavity frequency $\omega_\mathrm{c}$. This expression is valid for an arbitrarily high-$Q$ cavity with the lifetime $\tau_\mathrm{c}\to \infty$, and the resonant behavior is apparently enforced by the Gaussian function, where the maximum of the rate is reached when $\omega_\mathrm{c}=\omega_{0}$. Note that under this limit, the rate constant profile is purely controlled by the broadening $\sigma$ (see Eq.~\ref{sigma}), which is related to the phonon spectral density $J_{\nu}(\omega)$. In general there is still the broadening effect due to the $\tau^{-1}_\mathrm{c}$ term in $J_\mathrm{eff}$. To numerically show the effects of cavity lifetime, we set the phonon bath of $\hat{H}_\nu$ as the Drude-Lorentz form, 
$J_\nu (\omega) = 2\lambda \gamma \omega / (\omega^2 + \gamma^2)$,
where $\lambda$ is the reorganization energy, $\gamma$ is the characteristic frequency. Here we take $\gamma = 200~\mathrm{cm}^{-1}$, and $\lambda = 0.1 \omega_\mathrm{b} \gamma / 2$ (the same set up as Ref.~\citenum{Arkajit_2022}). The numerical value of the broadening factor at $T=300~\mathrm{K}$ is $\sigma=62.7$ cm$^{-1}$.

Second, under the limit when the phonon broadening is much smaller than the cavity caused broadening, which happens when $\tau^{-1}_\mathrm{c}\gg \sigma$ (for example, when $\tau_\mathrm{c}\to0$), the Gaussian function $G$ is then much narrower than the $J_\mathrm{eff}$, such that we can approximate $G(\omega-\omega_{0})$ as a single $\delta$-function, $G(\omega-\omega_{0})\approx\delta(\omega-\omega_{0})$, then the $k_\mathrm{VSC}$ expression in Eq.~\ref{kvsc-int} becomes
\begin{equation}\label{eq:FRR-approx}
    \tilde{k}_\mathrm{VSC}\approx k_\mathrm{VSC}(\omega_{0}),
\end{equation}
such that the cavity-related width $\tau^{-1}_\mathrm{c}$ dominates the rate profile. Under this limit, the rate reduces to the expressions presented in the main text: Eq.~15, or under the special cases, Eq.~17 for the fully aligned dipole and Eq.~18 for the isotropically distributed dipole.

\section{VII. Details of the Numerical Results in Fig. 1}
We provide details of the numerical results presented in Fig. 1 of the main text, to illustrate the VSC resonance phenomena, and how $k_\mathrm{VSC}$ is influenced by light-matter coupling strength (Rabi-splitting $\Omega_\mathrm{R}$) and cavity lifetime $\tau_\mathrm{c}$. The model parameters are provided in Table~\ref{tab_par} in Sec.~II. To simplify our argument, the VSC rate constant change $k_\mathrm{VSC}$ is calculated using Eq.~17 of the main text for the {\it fully aligned case} because we can directly relate the coupling strength with the Rabi splitting. For the fully isotropic case of the dipole (Eq.~18 of the main text), the rate constant will be 1/3 of the results shown here if the same $\sqrt{N}g_\mathrm{c}$ is used.

Fig.~1a presents $k_\mathrm{VSC}$ as a function of $\omega_\mathrm{c}$. The rate profile peaks at $\omega_\mathrm{c}=\omega_{0}$. As one increases the light-matter coupling strength $\Omega_\mathrm{R}$, the magnitude of the change also increases, but they all have a similar sharp-resonance profile. Fig.~1b presents the value of $k/k_{0}=1+k_\mathrm{VSC}/k_{0}$ at $\omega_\mathrm{c}=\omega_{0}$, as a function of $\Omega_\mathrm{R}$ (red curve), reading with the y-axis on the right-hand side. The value of $k_{0}$ needs to be computed separately, where we use $k_0 = 2.3 \times 10^{-6}~ \mathrm{fs}^{-1}$, a typical value obtained from the previous exact quantum dynamics simulations ~\cite{Arkajit_2022} for the same  ground-state reaction model investigated here. The overall scaling is quadratic with Rabi splitting, 
\begin{equation}\label{eq:predict1}
    {k}/{k_{0}}\propto 1+\mathcal{C}\cdot\left({\Omega_\mathrm{R}}/{2\omega_\mathrm{c}}\right)^2,
\end{equation}
where $\mathcal{C}$ is a constant factor. 

Fig.~1b further presents the effective change of the free energy barrier $\Delta(\Delta G^{\ddagger})$ (blue), directly backed out from the rate constants ratio $k/k_{0}$. To account for the ``effective change" of the Gibbs free energy barrier $\Delta(
\Delta G^\ddagger)$, we consider the simple rate equation~\cite{Ebbesen_nanophotonics_2020} $k = A\cdot \exp[-{\Delta G^\ddagger}/{k_\mathrm{B}T}]$, and outside the cavity case as $k_{0} = A\cdot \exp[-{\Delta G_{0}^\ddagger}/{k_\mathrm{B}T}]$, as commonly assumed by experimental analysis~\cite{Ebbesen_angew_2019,Ebbesen_nanophotonics_2020}. The pre-factor $A$ is assumed to be the same with or without the cavity. The change of free energy barrier compared to the bare molecular reaction (with  $k_0$ and $\Delta G^{\ddagger}_0$) is then 
\begin{equation}
    \Delta(\Delta G^\ddagger)=\Delta G^\ddagger-\Delta G^{\ddagger}_0=-k_\mathrm{B}T\ln \left({k}/{k_0}\right).
\end{equation}
Note that this is not an actual free energy barrier change, but rather a purely kinetic effect (as we best understood, through the FGR in Eq.~15, Eq.~17, and Eq.~18 of the main text). Based on our observation in Eq.~\ref{eq:predict1}, we {\it predict} that 
\begin{equation}\label{predit2}
    \Delta(\Delta G^\ddagger)\propto- k_\mathrm{B}T\ln \left(1+\mathcal{C}\cdot({\Omega_\mathrm{R}}/{2\omega_\mathrm{c}})^2\right).
\end{equation}
Note that if one hypothesizes that an unknown mechanism forces the upper or lower vibrational polariton states to be a ``gateway of VSC polaritonic" chemical reaction~\cite{Hiura2019-rate}, then the activation free energy change should shift linearly~\cite{TaoJCP2020} with $\Omega_\mathrm{R}$. The experimental results, on the other hand, demonstrate a non-linearity of reaction barrier~\cite{Ebbesen_nanophotonics_2020,Ebbesen_angew_2019}. Our current theory thus predicts a non-linear increase of the ``effective" $\Delta(\Delta G^{\ddagger})$ in Eq.~\ref{predit2} as increasing $\Omega_\mathrm{R}$ due the cavity promotion of the $|\nu_\mathrm{L}\rangle\to|\nu'_\mathrm{L}\rangle$ transition, and more particularly, the effective $\Delta(\Delta G^\ddagger)$ scales with $-2k_\mathrm{B}T\ln \Omega_\mathrm{R}$. Further, in Ref.~\citenum{Hirai2020C}, it was pointed out that for a very small Rabi splitting ($\Omega_\mathrm{R}=100$ cm$^{-1}$) observed in optical spectra, it can lead to much larger changes in activation free energy $\Delta(\Delta G^{\ddagger})\approx0.8$~kcal/mol or 3.3 kJ/mol), such that $\Delta(\Delta G^{\ddagger})>\Omega_\mathrm{R}$. This seems to be a general trend in most VSC experiments~\cite{Ebbesen_nanophotonics_2020} but this phenomenon lacks a theoretical explanation. Here, we provide one due to the FGR nature of the rate constant, which significantly influences the rate and correspondingly, the effective reaction-free energy barrier.

\begin{figure*}[htbp]
\centering
\includegraphics[width=0.7\linewidth]{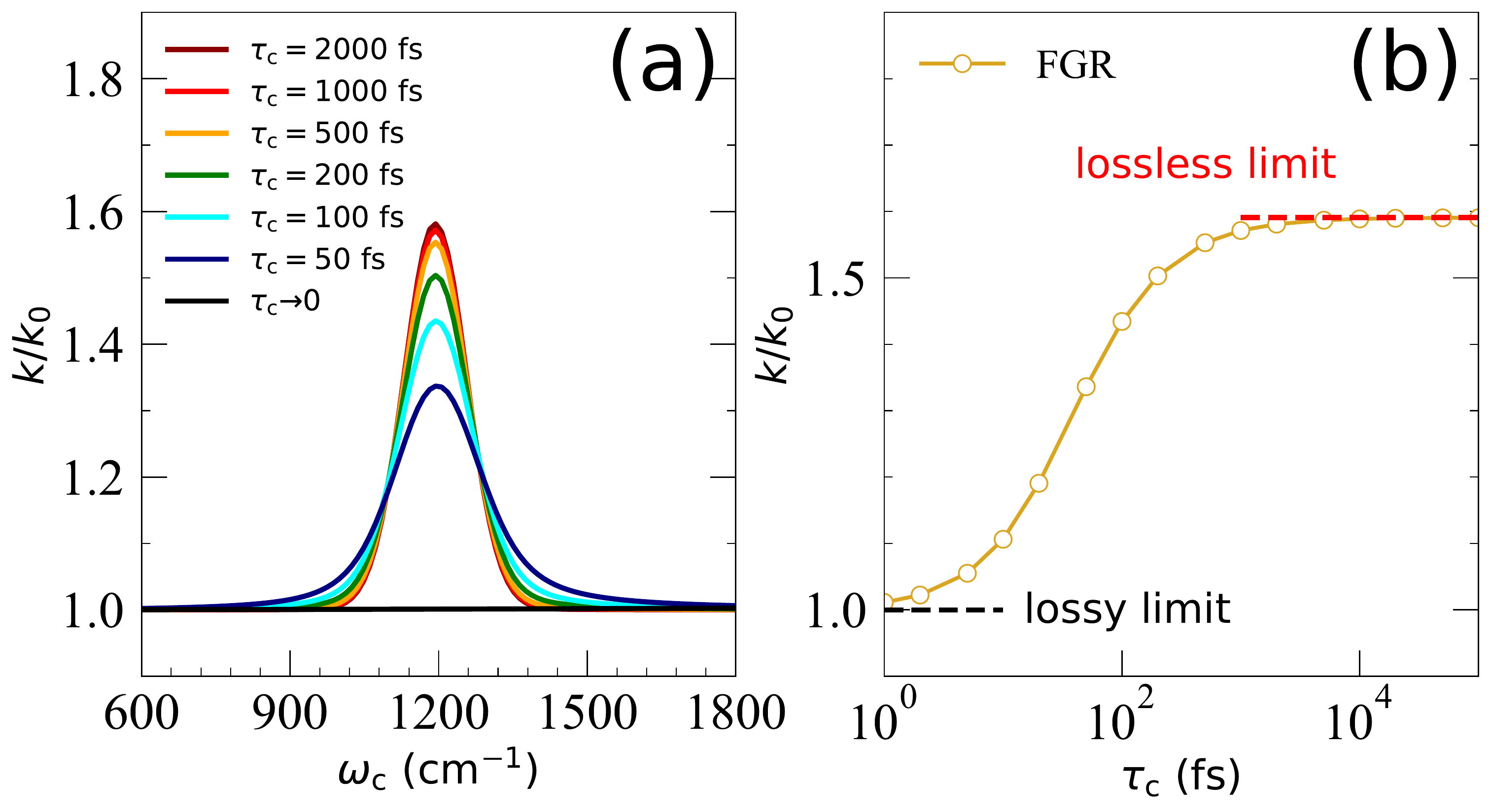}
\caption{(a) Effect of $\tau_\mathrm{c}$ on the rate constant ratio $k/k_{0}$. Note that the cavity modification effects become smaller when $\tau_\mathrm{c}$ is reduced and under the heavy loss limit ($\tau_\mathrm{c} \rightarrow 0$) the cavity effect vanishes. (b) The peak value of $k/k_{0}$ (at $\omega_\mathrm{c}=\omega_{0}$ as a function of cavity life time $\tau_\mathrm{c}$. The Rabi-splitting is fixed at $\hbar\Omega_\mathrm{R} = 100~\mathrm{cm}^{-1}$. }
    \label{fig_tauc}
\end{figure*}

Fig.~\ref{fig_tauc}a presents $\tilde{k}_\mathrm{VSC}$ as a function of cavity frequency $\omega_\mathrm{c}$, as well as cavity lifetime $\tau_\mathrm{c}$, using a numerical evaluation of Eq.~\ref{kvsc_lossless} and the model spectral density $J_{\nu}(\omega)$ mentioned in section VI.  Decreasing $\tau_\mathrm{c}$ leads to a weakening and broadening of the resonance rate peak. The cavity modification effect gradually disappears at the heavy loss limit $\tau_\mathrm{c} \rightarrow 0$. To the best of our knowledge, there is no existing experiment that directly verifies the $\tau_\mathrm{c}$-dependence of the VSC effect. Future experimental efforts could be dedicated to this specific check. Fig.~\ref{fig_tauc}b, which is the identical plot of Fig. 1c in the main text, presents the value of $k/k_{0}$ at $\omega_\mathrm{c}=\omega_{0}$, as a function of $\tau_\mathrm{c}$ (on a log scale in $\tau_\mathrm{c}$). The rate constant modification changes in a sigmoid trend, which is subject to future experimental verification. 

\providecommand{\noopsort}[1]{}\providecommand{\singleletter}[1]{#1}%